\documentclass[preprint,5p,twocolumn]{elsarticle}

\biboptions{sort&compress}
\journal{Applied Energy}
\usepackage{mathtools}
\usepackage{optidef}
\usepackage{amsmath}
\usepackage{amssymb}
\usepackage{algpseudocode}
\usepackage{algorithm}
\usepackage{bm}
\usepackage[font=small,labelfont=bf,tableposition=top]{caption}
\usepackage{eurosym}
\DeclareRobustCommand{\officialeuro}{%
  \ifmmode\expandafter\text\fi
  {\fontencoding{U}\fontfamily{eurosym}\selectfont e}}
\usepackage{gensymb}
\usepackage[utf8]{inputenc}
\usepackage[hidelinks]{hyperref}
\usepackage{multirow}
\usepackage{overpic}
\usepackage{pgfplots}
\usepackage{rotating}
\usepackage{siunitx}
\usepackage[font=footnotesize]{subfig}
\usepackage{tikz}
\usepackage{xcolor}

\usetikzlibrary{calc}
\usepgfplotslibrary{units}
\pgfplotsset{unit code/.code={\si{#1}}}
\newtheorem{remark}{Remark}

\newcommand{\rl}[1]{_\mathrm{#1}}
\newcommand{\tr}{^{\top}}
\newcommand{\rs}[1]{_\mathrm{#1}}
\newcommand{\ru}[1]{^\mathrm{#1}}

\newcommand{\blue}[1]{\textcolor{black}{#1}}
\newcommand{\white}[1]{\textcolor{white}{#1}}

\newcolumntype{C}[1]{>{\centering\let\newline\\\arraybackslash\hspace{0pt}}m{#1}}
\newlength\figH
\newlength\figW

\begin{document}

\begin{frontmatter}

\title{Scenario-based Nonlinear Model Predictive Control for Building Heating Systems}

\author[a,1]{Tomas Pippia\corref{cor1}}
\ead{t.m.pippia@tudelft.nl}
\author[a,1,2]{Jesus Lago}
\author[3]{Roel De Coninck}
\author[1]{Bart De Schutter}

\cortext[cor1]{Corresponding author}
\address[1]{Delft Center for Systems and Control, Delft University of Technology,
	Mekelweg 2, Delft, The Netherlands}
\address[2]{Algorithms, Modeling, and Optimization, VITO, Energyville,
	ThorPark, Genk, Belgium}
\address[3]{DeltaQ, Quai à la Chaux 6, Brussels, Belgium\\\vspace{0.4cm}$^1$These authors contributed equally to this work}

\begin{abstract}
State-of-the-art Model Predictive Control (MPC) applications for building heating adopt either a deterministic controller together with a nonlinear model or a linearized model with a stochastic MPC controller. 
However, deterministic MPC only considers one single realization of the disturbances and its performance strongly depends on the quality of the forecast of the disturbances, \blue{which can lead to low performance. In fact, inadequate building energy management can lead to high energy costs and CO$_2$ emissions}. On the other hand, a linearized model can fail to capture some dynamics and behavior of the building under control. In this article, we combine a stochastic scenario-based MPC (SBMPC) controller together with a nonlinear Modelica model that is able to \blue{provide a richer building description and to} capture the dynamics of the building more accurately than linear models. 
The adopted SBMPC controller considers multiple realizations of the external disturbances obtained through a statistically accurate model, so as to consider different possible disturbance evolutions and to robustify the control action. \blue{To this purpose, we present a scenario generation method for building temperature control that can be applied to several exogenous perturbartions, e.g.\ solar irradiance, outside temperature, and that satisfies several important stastistical properties, in contrast with simpler and less accurate methods adopted in the literature.}
We show the benefits of our proposed approach through several simulations in which we compare our method against the standard ones from the literature, \blue{for several combinations of a trade-off parameter between comfort and energy cost}. We show how our \blue{SBMPC controller approach outperforms the standard controllers available in the literature}.
\end{abstract}

\begin{keyword}
Scenario-based model predictive control \sep Control of Buildings \sep Model predictive control \sep Modelica 
\end{keyword}

\end{frontmatter}

\section{Introduction}\label{sec:Intro}
Energy consumed in buildings for heating, ventilation, and air-conditioning (HVAC) purposes accounts for around half of total energy used in buildings \cite{YanYan:14,LiXua:20,eurostat_building,Eurostat2018,Lago2019}. For companies, especially if they are located in large buildings, it is therefore very important to limit the amount of energy wasted due to bad temperature control. Furthermore, it is also important to reduce as much as possible the energy waste in order to decrease emissions due to e.g.\ natural gas boilers \cite{BloSch:18}. Moreover, buildings have comfort temperature bounds that have to be respected during working hours, with few violations allowed \cite{IndoorParameters}. The comfort violations should be limited while at the same time the energy cost has to be minimized. Simple solutions where heaters always run at maximum power are not acceptable due to the large costs and energy waste.

On top of that, buildings are subject to many exogenous disturbances, e.g.\ outside temperature, solar irradiance, and endogenous ones, e.g.\ occupancy. The profile of these disturbances, if not properly considered when determining the control actions, can lead to poor performance, both in terms of energy cost and discomfort. Moreover, the model of the building considered, due to approximations, may lead to further errors. However, a too complex model is also not useful for control purposes due to the high computational burden that it entails. Therefore, the problem of controlling the room temperature in large buildings is a complex task.

\subsection{Literature review}
In this section, we perform a brief literature review of the two main topics of this research: control for buildings and scenario generation.

\subsubsection{Modeling and control strategies for buildings}

Many solutions have been proposed in the literature to control the room temperature in buildings using information available on the current temperature and forecasts of the disturbances. A simple solution involves a rule-based approach that is based on if-then-else rules and information about the current disturbances \cite{AghVir:13}. Although these controllers are simple to implement and may achieve a reasonable performance, they are not very efficient since they are based on user knowledge and rules-of-thumb and they do not actually perform an optimization. Model predictive control (MPC) \cite{CamAlb:13,Mac:02,May:14,FarGiu:16} is a more advanced control strategy that is suitable for the room temperature control problem, since it can naturally include constraints in the control problem and since it can in general achieve a better performance \cite{DeCHel:16,DrgPic:18,OldPar:12}. 
Moreover, a building is subject to several disturbances, as mentioned before, and MPC can deal well with disturbances by using a robust or stochastic controller \cite{May:14}, which can achieve better performance than the deterministic counterpart. However, despite having a better constraint satisfaction, a robust MPC controller \blue{for buildings, e.g.\ \cite{OstDub:20},} could be too conservative for the task of controlling the temperature of a room and it would result in a high amount of energy used as it would try to compensate every possible disturbance realization. Therefore, usually a stochastic MPC controller is preferred for building heating systems \cite{FarGiu:16,ParMol:13}. Indeed, by considering the stochastic properties of the disturbance or by considering several disturbance scenarios, stochastic MPC controllers can potentially achieve a better control performance compared to deterministic controllers, leading therefore to a reduced energy consumption while limiting the discomfort. In particular, scenario-based MPC (SBMPC) methods stand out as a useful tool in building heating systems, since they can use past data of the disturbances, which are available in the case of building heating systems, and they can successfully be applied to nonlinear models as well \cite{FarGiu:16}.

In this regard, several types of MPC algorithms have been applied to HVAC systems in the literature \cite{ParMol:13,OldPar:12,ParVar:13,DeCHel:16,ParFab:14,LonLiu:14,TanWan:19,KubHeb:19,AouBav:19,DrgPic:18,JorBoy:19,Zhang2013,CarCav:20,PipLag:19,LiZha:21,YasVec:21}; see also \cite{WenMis:18,MadLia:20} and the references therein. In particular, \cite{OldPar:12} presents two stochastic MPC algorithms, i.e.\ a disturbance-feedback approach and a chance-constraint one. The results show that the  stochastic controllers achieve a better performance than deterministic MPC and rule-based control. Authors of  \cite{ParMol:13} develop an SBMPC controller \blue{that uses previous data of building occupancy and external ambient conditions forecast errors to solve a scenario-based optimal control problem. The scenarios are built through copulas that can be learned online and the method is applied to a room in a university building. The results show that the proposed controller is able to achieve a good performance in terms of energy cost, while having a larger computational complexity than standard deterministic methods. However, the method is applied to a single room and the authors of \cite{ParMol:13} suggest that a study needs to be carried out to asses whether the method can be applied to a larger building.} A similar approach is presented in \cite{ParVar:13}, \blue{where the main differences are that 1) slack variables are introduced in the cost function to improve the feasibility of the optimization problem obtained and 2)  that different copula families are tested and compared (see also Section 1.1.2). In both papers, the authors do not rely on Gaussian assumptions for what concerns the properties of the disturbances. Given that the two previous methods might result in a large computational complexity, the authors of \cite{ParFab:14} extend the} concept to an explicit SBMPC controller, \blue{such that the control inputs are computed offline and applied online. Experiments were performed once again for a single university room, showing better performance with respect to the standard methods. In order to deal with a multi-room setting, a distributed MPC controller is presented in \cite{LonLiu:14}, where a Lagrangian dual decomposition relaxation method is used to reduce the computational burden arising from the several rooms considered. Simulation results obtained considering a network of 10 rooms show an increased performance with respect to a baseline PID controller}. In all these articles it is shown how stochastic MPC strategies can achieve a better performance than deterministic MPC. The article \cite{TanWan:19} presents an MPC algorithm in which a linear model is used to control a building including an active cold thermal storage in order to implement a demand response program. All these works, i.e.\ \cite{OldPar:12,ParMol:13,ParVar:13,ParFab:14,LonLiu:14,TanWan:19}, use a linearized model description and do not use a nonlinear model nor other \blue{more advanced} modeling tools, e.g.\ Modelica \cite{Modelica,Fri:15}, \blue{possibly leading to a decrease in the performance}. Such tools and nonlinear models for building heating control \blue{usually include more features and components compared to a linear model and thus} they can \blue{potentially} provide a more accurate description of the building and of the influence of each disturbance, reducing therefore the modeling error and improving the overall performance. The article \cite{KubHeb:19} adopts a nonlinear model arising from the heat pump and a battery inverter considered, but the considered MPC controller is a deterministic one.
For what concerns nonlinear \blue{and more advanced} models, while some works did consider their usage for HVAC systems, e.g.\ \cite{JorBoy:19,DrgPic:18,DeCHel:16,AouBav:19}, all of them considered a deterministic setting instead of a stochastic one. To the best of our knowledge, no work has considered a nonlinear model description obtained through Modelica \textit{together with} a stochastic controller, which would improve the performance by taking into account a more accurate model and  the stochastic properties of the disturbances. 

\subsubsection{Scenario generation}
\label{sec:introscen}
Besides improving the state-of-the-art by proposing a control approach for more realistic models, i.e.\ nonlinear Modelica models, our work also contributes to the existing literature of scenario generation for buildings by improving upon the current state-of-the-art. In particular, scenarios of random variables that represent a time series, e.g.\ the ambient temperature for the next 24 hours with an hourly resolution, need to satisfy several important properties: 

\begin{enumerate}
	\item They should not be restricted to the standard assumption of Gaussian disturbances/forecasting errors as this assumption is quite restrictive when it comes to generating scenarios of heteroscedastic\footnote{A time series variable is heteroscedastic if the variance changes throughout time.} processes, e.g.\ solar irradiance \cite{ParFab:14}.
	\item \label{ls:prop1} They need to consider the multivariate distribution of the random variables: if the scenarios represent a random variable at different time steps in the future, these scenarios should model the time correlation of the random variable \cite{PinMad:08}.
	\item \label{ls:prop2} Besides modeling the time correlation, they should explicitly take into account the different time dependencies and avoid modeling a stationary distribution; i.e.\ the distribution of the random variable might be different at different hours of the day/times of the year or might change with the prediction horizon.
	\item \label{ls:prop3} The methods to generate scenarios should be flexible enough to explicitly model the dependencies of the random variables on exogenous variables.
	\item \label{ls:prop4} The computational burden of the scenario generation method should be small enough for online implementation.
\end{enumerate}

In the context of building heating, while some scenario generation methods have been considered \cite{Ma2012,Ma2012a,OldPar:12,ParMol:13,ParVar:13,LonLiu:14,ParFab:14,Pedersen2018,Deori2014,Zhang2013,Maiworm2015,Hedegaard2017}, they have several problems. In particular, some of the existing methods \cite{Ma2012,Ma2012a,OldPar:12} rely on the standard Gaussian assumptions \cite{ParFab:14}. In addition, although several works have addressed the Gaussian assumption \cite{ParMol:13,ParVar:13,LonLiu:14,ParFab:14,Pedersen2018,Zhang2013,Maiworm2015,Hedegaard2017}, they still lack some of the required properties.

 More specifically, in \cite{ParMol:13}, a method based on empirical copulas is proposed. While the method satisfies some properties, e.g.\ time correlation, it fails to satisfy the following two: 1) it does not model time dependencies but it assumes that the marginal distributions are stationary, i.e.\ it assumes that the $n$-hours ahead distribution of a variable is the same at any hour of the day, any day of the year; 2) the scenarios are generated based on historical data without considering other possible exogenous inputs. The analytical copula method proposed in \cite{LonLiu:14} overcomes some of these issues as it explicitly models the time dependency during a day. However, the distributions are still stationary, i.e.\ they vary within a day but they do not change along a year, and they are just based on historical data. In \cite{ParVar:13} and \cite{ParFab:14}, a more general approach is proposed where different copula families are tested, and the best one is selected to generate scenarios. While the method is very general and flexible, it requires to compare different copula functions and can easily become computationally infeasible for online MPC. In addition, the method has two other problems: the best copula is selected by comparison with the empirical copula of \cite{ParMol:13}, hence it has the same problems as \cite{ParMol:13}; moreover, the time dependencies considered by the copulas are not specified. In \cite{Pedersen2018}, scenarios from a weather meteorological model are employed. Even though the goal of weather models is to provide an ensemble of scenarios, to capture the uncertainty in the prediction, the method displays systematic errors, e.g.\ biases, and requires the application of advanced post-processing techniques based on copulas, e.g.\ ensemble copula coupling \cite{Schefzik2013}. In \cite{Zhang2013}, a method based on sampling historical forecasting errors is considered. Despite the method attempts to capture time correlation using an auto-regressive error model, that model is only used for error correction. In particular, to generate scenarios, the method samples from past historical errors and fails to satisfy properties \ref{ls:prop1}-\ref{ls:prop3}. The recursive feasibility and stability of SBMPC is studied in \cite{Maiworm2015}. To do so, it is assumed that disturbances can be represented by a scenario tree, and that the tree can be built using empirical samples from a discrete set of disturbances. This approach is clearly very limited as, it does not satisfy properties \ref{ls:prop1}-\ref{ls:prop3}, and in addition it could have scalability  and computational issues when the number of random variables increases. In \cite{Hedegaard2017}, scenarios are used for modeling electricity prices and independent optimization problems are solved for each scenario; however, the method cannot be used to model uncertainty in other variables, e.g.\ weather variables, and fails to satisfy properties \ref{ls:prop1}-\ref{ls:prop3}. In \cite{Deori2014}, two Poisson distributions are employed to model the occupancy in the building as a birth-death process. While such a parametric distribution might work well for occupancy, it has the same issue as the Gaussian assumption: the method cannot be generalized to other random variables.


\subsection{Motivation and contributions of the paper}
In this article, we focus on a scenario-based MPC (SBMPC) algorithm that includes both a nonlinear system description through Modelica, while using a scenario generation method based on probability distributions obtained empirically, making it a very suitable tool for a building heating control problem. The Modelica model description \blue{can lead to an improvement of} the model accuracy; \blue{note that in} the current literature of SBMPC for heating systems in buildings, a linearized model is always used. On top of that, by using a nonlinear \blue{Modelica} model, we \blue{implement for the first time an SBMPC controller} for HVAC systems in buildings \blue{that uses a nonlinear Modelica model description}. In addition, we \blue{present} a parametric Gaussian copula method to generate scenarios that it can satisfy the five required properties, \blue{whereas the methods used in the literature of HVAC systems for scenario generation suffer from some statistical problems, as mentioned earlier}. \blue{We perform several simulations showing the benefits of the presented approach}. Our contribution is therefore threefold:

\begin{itemize}
\item We propose a control method for a building heating systems that considers a Modelica nonlinear model and an SBMPC controller. 
\item We generate scenarios using a new approach that, unlike the existing methods, satisfies all the important properties of scenario generation methods for time \mbox{series} data.
\item We perform a comparison between several combinations of the couple model-controller: as models, we consider a Modelica and linear and as controller we consider a deterministic MPC and SBMPC.
\end{itemize} 

\subsection{Outline}
The outline of the article is as follows. In Section \ref{sec:Model}, we describe the problem under consideration. We present the control algorithms used throughout the article in Section \ref{sec:Control}. In Section \ref{sec:Scenarios}, the adopted scenario generation method is presented. We present the simulation results on a case study in Section \ref{sec:CaseStudy} and lastly we present some conclusions and suggestions for future work in Section \ref{sec:Conclusions}.

\section{Model description}\label{sec:Model}
\subsection{Buildings}
\label{sec:officebuilding}
In this paper, we focus our attention on buildings with local heat production units. The type of building considered can be controlled via two control inputs, i.e.\ $\bm{u}~=~\begin{bmatrix}
q\ru{heat} & q\ru{cool}\end{bmatrix}^\top$, where $q\ru{heat}$ is the amount of heating power transferred to the building and $q\ru{cool}$ is the cooling power provided to the building. We assume that the building can be modeled using an RC-model with two states \cite{DeCHel:16}: $T\ru{zone}$ as the average temperature of the rooms and $T\ru{wall}$ as the average temperature of the walls. In addition, it is assumed that the building is affected by three disturbances: solar irradiance $I$, outdoor temperature $T\ru{amb}$, and building occupancy $\theta\ru{occ}$. While past measurements of external disturbances, e.g.\ solar irradiance and outdoor temperature, are available, we do not have any measurement of the occupancy of the building. Note that, although this is an important disturbance to consider, it is also difficult to measure in practice \cite{LabZei:15,SolWhi:18}. Therefore, to estimate the models and to perform simulations, we assume that the occupancy profile is fixed for every day of the week, i.e.\ we assume that the building is fully occupied during working hours and empty outside of these hours.

\subsection{Modelica}
We have modeled the buildings, comprising also the heating, cooling, and ventilation units, with Modelica \cite{Modelica,Fri:15}, which is an object-oriented and equation-oriented language that is designed to model the behavior of physical systems.  In particular, the building is modeled based on an RC-model, which has been identified through the Grey-Box Buildings toolbox \cite{DeCMag:16}. The building has also been extensively validated using data collected from the building as in \cite{DeCMag:16,DeCHel:16}. \blue{Such data has been gathered between 2016 and 2018 and includes e.g.\ internal temperature, domestic hot water usage, external temperature, and solar irradiance.} The adoption of Modelica in our work provides the benefit that we can improve the amount of detail and accuracy of the model w.r.t.\ linear models. Note that e.g. some of the HVAC components modeled in Modelica result in nonlinear model components. Therefore, although many other works, e.g.\ \cite{ParMol:13,OldPar:12,LonLiu:14,ParVar:13,ParFab:14}, use indeed a linearization of a nonlinear model, in this work we directly use a nonlinear model and we obtain therefore a more meaningful representation of the real building. Readers interested in the modeling procedure of buildings in a Modelica environment are referred to \cite{DeCHel:16,DeCMag:16}.

Note that other high-fidelity simulation tools exist for buildings, e.g.\ TRNSYS, EnergyPlus, ESP-r, IDA ICE; see \cite{CraHan:08,ShaBin:14} for a complete review. However, compared to Modelica, these software tools lack in modularity and flexibility for prototyping and simulating new technologies \cite{AndSae:18}. Moreover, \blue{in our particular setting, we have data available from the buildings to be controlled that allows us to estimate the parameters of the building. However, the data is not comprehensive enough for satisfactory white-box modeling. Therefore, as highlighted in \cite{DeCMag:16}, we adopt a grey-box model due to necessity of estimating certain parameters of the model of the building that are not known a priori. Whereas in the aforementioned white-box modeling tools it could in theory be possible to calibrate the parameters of the model, the lack of a more detailed physical knowledge of the building makes the choice of a white-box model, for our specific setting, less desirable.} Lastly, note that Modelica is an open-source tool, making it particularly appealing for commercial applications. For more advantages of using Modelica for HVAC systems we refer to \cite{Wet:09}.

\begin{remark}
Note that, as mentioned in \cite{DeCMag:16}, the actual difference between white-box and grey-box models does not depend on the complexity of the model. For instance, even a single-state model can be a white-box model if all its parameters can be determined based solely on physical knowledge. However, a white-box model becomes grey when one or more of its parameters are estimated based on a fitting of the model to measurement data, irrespectively of the complexity of the white-box model.
\end{remark}

\begin{remark}
The model of the building used in this paper has been extensively validated following the exact same procedure reported in \cite{DeCMag:16}. Therefore, the validation process is omitted here as it is already explained in the aforementioned paper. Multiple-years data has been used to identify and validate the model. The interested readers might find the whole identification and validation procedure in \cite{DeCMag:16,DeCHel:16}.
\end{remark}

\subsection{Model predictive control}\label{sec:MPC_background}
MPC is a control tool that has been extensively studied in the last forty years \cite{Mac:02,CamAlb:13,May:14}. The main strength of MPC is to use a model of the system under control to find optimal inputs for that system with respect to a certain objective, which can be either stabilizing the system or minimizing an economical goal. Since in the MPC framework the problem of finding the optimal inputs for the system is an optimization problem, constraints can easily be included, as well as several performance criteria. Compared to simpler strategies e.g.\ rule-based control, MPC requires a larger computational effort and requires to solve an optimization problem online but it can provide an increase in performance. \blue{Due to its versatility and good performance achieved, MPC has been applied to various systems and fields, e.g.\ power systems \cite{PipSij:19,KorKho:13}, traffic networks \cite{LiuHel:17}, water networks \cite{KorSaf:16}, aerospace \cite{ErePra:17}, among others. For a survey that includes also MPC applications for large-scale and industrial systems, we refer the reader to \cite{KorSaf:21}.}

In recent years, several MPC strategies have been developed to cope with external disturbances, in particular, robust and stochastic MPC strategies. In this paper we focus on an SBMPC algorithm, which is presented in Section \ref{sec:SBMPC}.

\subsection{Control loop and practical implementation}\label{sec:implementation}
\begin{figure}[t]
	\vspace{0.2cm}
	\centering
	\includegraphics[scale=0.80]{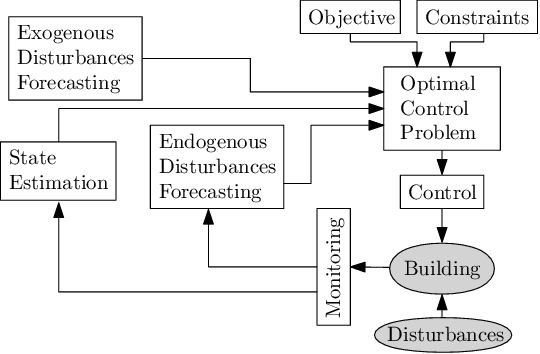}
	\caption{Scheme of the MPC framework (adapted from \cite{DeCHel:16}).
	}
	\label{fig:overall_scheme}
\end{figure}

Many operations have to be carried out on the real building by the building energy management system; the overall control scheme is presented in Figure \ref{fig:overall_scheme}. The operations are \cite{DeCHel:16}:

\begin{enumerate}
\item Monitoring: some measurements, e.g.\ water temperature, heat flux, are performed by the building energy control and management system.
\item Forecasting: weather forecasts are obtained as will be explained in Section \ref{sec:forecasting}.
\item State estimation: some states, e.g.\ internal wall temperatures, cannot be measured and therefore they have to be estimated.
\item Optimization of the control inputs: an optimization problem, explained in Section \ref{sec:Control}, is solved at every time step with a sampling time of 1$\mathrm{h}$. Only the first inputs of the optimal sequence are applied to the system. 
\end{enumerate}

The real building is therefore controlled by all these steps, carried out in sequence. 
The optimization problem discussed in step 4 above is solved through JModelica.org \cite{AkeArz:10}. The direct collocation method is used to discretize time so that the optimization problem is reduced to a nonlinear programming problem \cite{MagAke:12}. CasADi \cite{AndGil:18} is used to obtain the first-order and second-order derivatives of the expressions in the nonlinear programming problem with respect to the decision variables, which are required by the solvers used by JModelica.org. We use IPOPT \cite{WacBie:06} to solve the nonlinear programming problem, together with the sparse linear solver MA57 \cite{hsl}.

\begin{remark}\label{rem:simulationScheme}
Note that in this section we consider simulations instead of experiments in the real building. This implies a small difference w.r.t.\ Figure \ref{fig:overall_scheme}: instead of applying the inputs to the real building, we simulate its behavior through Modelica for one time step. In other words, the loop is ``closed'' by applying the optimal inputs to a model of the building rather than to the building itself, using the actual values of the disturbances instead of the forecasts. The model used for simulating the building behavior is the same Modelica model used for the optimal control problem in the MPC framework.
\end{remark}

\subsection{Linear model estimation}
\label{sec:linearization}
To compare the nonlinear MPC controller with the standard linear counterpart, a linear model of the building is needed. To obtain such a model, data from the building is considered and a linear model is estimated using linear least squares. In detail, considering the same inputs, state space, and disturbances as for the nonlinear model (see Section \ref{sec:officebuilding}), we can assume that the building dynamics are of the form:

\begin{equation}
\label{eq:linearmodel}
    \begin{bmatrix} T\ru{zon}_{k+1}\\ T\ru{wall}_{k+1} \end{bmatrix} = A  \begin{bmatrix} T\ru{zon}_{k}\\ T\ru{wall}_{k} \end{bmatrix} + B_1  \begin{bmatrix} q\ru{heat}_{k} \\ q\ru{cool}_{k}  \end{bmatrix}+ B_2 \begin{bmatrix} T\ru{amb}_{k} \\ I_{k} \\\theta\ru{occ}_{k}  \end{bmatrix}.
\end{equation}

Then, using the same data as those used for estimating the Modelica nonlinear model, we solve a linear least square problem and estimate the values of the matrices $A$, $B_1$, and $B_2$, using the mean absolute error as key performance indicator.

\blue{
\begin{remark}\label{rem:linearModel}
In this article, as highlighted in Remark \ref{rem:simulationScheme}, we use the nonlinear Modelica model to compute the next state for the closed-loop simulations. This can introduce some bias when comparing the performance of the controllers that use the linear model with respect to the ones that use the Modelica model, since there would be no model mismatch in the latter case. Nevertheless, we present here also a linearized model as a reference, to show how such model would behave in the considered scenario. 
\end{remark}
\begin{remark}\label{rem:nonlinearModelTested}
Related to the previous remark, note also that this kind of nonlinear Modelica model has been already applied to a building heating system in \cite{DeCHel:16}. Indeed, in that reference the same modeling procedure was applied to a similar building and experimental results were carried out. No large mismatch between the Modelica nonlinear model and the actual physical building nor other fundamental flaws were noticed. Therefore, we can argue that the nonlinear Modelica model is a very good approximation of the real building and can therefore be used in the closed-loop simulations to simulate the evolution of the system.
\end{remark}
}

\section{Control algorithms and models}\label{sec:Control}
In this section we present the two considered control algorithms, i.e.\ MPC and SBMPC.
\subsection{Deterministic MPC} \label{sec:DetMPC}
In Deterministic MPC, the external disturbances, e.g.\ temperature or solar irradiance, are predicted with a point forecasting technique and in which the predictions are then assumed to represent the expected value. In this context, at each time step, the MPC optimization problem is solved, yielding an optimal control input sequence $\bm{u}^*$. Then the first element of the sequence is applied, the horizon is moved one time step forward, the system is sampled, and the optimization problem is solved again.

Given the task of controlling the room temperature in a building while minimizing both the energy costs and the discomfort, the optimization problem solved at each time step by a deterministic MPC controller is given by:

\begin{mini!}[3]{\substack{\displaystyle T_1,q_1\ldots,\\\displaystyle q_N, T_{N+1}}}{\sum_{k=1}^N \left(\alpha\, J\ru{d}_k + J\ru{e}_k\right) + \alpha\, J\ru{d}_{N+1} \label{eq:costDet}}{\label{eq:optidet}}{}
	\addConstraint{T_1}{=\overline{T}_1}
	\addConstraint{T_{k+1}}{= f(T_k, q_k, d_k),}{\quad \text{for }~k=1,\ldots,N \label{eq:dyn}}
	\addConstraint{0}{\leq q_k\ru{heat} \leq \overline{Q}\rs{max}\ru{heat},}{\quad \text{for }~k=1,\ldots,N}	
	\addConstraint{0}{\leq q_k\ru{cool} \leq \overline{Q}\rs{max}\ru{cool},}{\quad \text{for }~k=1,\ldots,N}
\end{mini!}

\noindent where:

\begin{itemize}
\item $N$ is the prediction horizon.
    \item The system state is defined by $T_k = [T_k\ru{zon}, T_k\ru{wall}]$, with $T_k\ru{zon}$ and $T_k\ru{wall}$ as the room and wall temperatures.
    \item $\overline{T}_1$ is the current temperature.
    \item The input control is defined by  $q_k = [q_k\ru{heat}, q_k\ru{cool}]$, with $q_k\ru{heat}$ and $q_k\ru{cool}$ as the input heating/cooling power.
    \item The cost function represents the weighted average between the energy cost $J\ru{e}_k$ and the discomfort cost $J\ru{d}_k$:

\begin{alignat}{3}
J\ru{d}_k & = \bigl(&&\max(T\ru{zon}_k - T\ru{max}_k,0) + \min(T\ru{zon}_k - T\ru{min}_k,0)\bigr)^2,\\
J\ru{e}_k & = c\ru{gas}_k \frac{q\ru{heat}_k}{\eta\ru{gas}_k} +  c\ru{ele}_k \frac{q\ru{cool}_k}{\eta\rs{cool}},\span\span
\end{alignat}

and $\alpha$ is the weighting parameter that defines the relative importance of each cost.
\item The building dynamics are defined by \eqref{eq:dyn}, where $ f(\cdot)$ represents the Modelica model of the building.
\item The building is disturbed by some uncontrollable inputs $d_k=[T\ru{amb}_k, I_k, \theta\ru{occ}_k]$, with $T\ru{amb}_k$ the ambient temperature,  $I_k$ the solar irradiance, and $\theta\ru{occ}_k$ the building occupancy.
\item  The upper and lower comfort temperature bounds are respectively defined by $T\ru{max}_k$ and $T\ru{min}_k$, and they vary in time depending on the hour of the day and day of the week.
\item $\overline{Q}\ru{heat}\rs{max}$, $\overline{Q}\ru{cool}\rs{max}$, $\eta\ru{cool}$, ${\eta}\ru{gas}$, $c\rs{gas}$, and $c\rs{ele}$ are constant parameters and are, respectively, the maximum heating power, the maximum cooling power, the cooling efficiency, the heating efficiency, the gas cost, and the electricity cost.
\end{itemize}

It is important to note that the role of the discomfort cost $J\rl{d}$ is to act as a soft constraint so that it penalizes the deviations of the temperature outside the comfort bounds, but remains $0$ if the temperature is inside the bounds. The controller can therefore choose to implement a control action that leads to a violation of the comfort bounds if this can lead to a lower total cost. \blue{Lastly, note that in \eqref{eq:costDet} the final cost is related only to the states. This is standard in the MPC framework (see e.g.\ \cite{May:14}, Eq.\ 2.3), since the inputs are applied from $k+1$ until $k+N$, thus the evolution of the system is considered from $k+2$ until $k+N+1$. Therefore, there is no input cost considered for time-step $k+N+1$, but there is a state cost that is the one we obtain by applying the inputs at $k+N$.}

\subsection{Scenario-based MPC}\label{sec:SBMPC}
It is possible to improve the performance of the deterministic MPC of the previous subsection by considering several scenarios of the disturbances acting into the system. This approach, known as scenario-based MPC (SBMPC), considers multiple realizations/scenarios for the disturbances, different system states for each scenario, and a cost function that consists of the average of the original cost functions across all scenarios. For the control inputs, two possibilities exist: different control inputs for each scenario (as with the system state) and shared control inputs across all scenarios. While the former has the advantage of being less conservative, the latter is more computational friendly. For the case of building control, we consider shared control inputs across all scenarios as this reduces the computational complexity.


Defining $M$ different scenarios for the disturbances. i.e.\ $d =\{\{d_{k,i}\}_{i=1}^M\}_{k=1}^N$, the SBMPC optimization problem solved at each time step can be defined as:

\begin{mini!}[3]
{\substack{\displaystyle T_1,q_1\ldots,\\\displaystyle q_N, T_{N+1}}}{\sum_{i=1}^M\Biggl(\sum_{k=1}^N \left(\alpha\, J\ru{d}_{k,i} + J\ru{e}_{k,i}\right) + \alpha\, J\ru{d}_{N+1,i}\Biggr)}{\label{eq:optisb}}{}
	\addConstraint{T_{1,i}}{=\overline{T}_1}{\quad \text{for }~i=1,\ldots,M}
	\addConstraint{T_{k+1,i}}{= f(T_{k,i}, q_k, d_{k,i}),}{\quad \text{for }~i=1,\ldots,M \nonumber}
	\addConstraint{}{}{\quad \text{for }~k=1,\ldots,N \label{eq:dyn2}}
	\addConstraint{0}{\leq q_k\ru{heat} \leq \overline{Q}\rs{max}\ru{heat},}{\quad \text{for }~k=1,\ldots,N}	
	\addConstraint{0}{\leq q_k\ru{cool} \leq \overline{Q}\rs{max}\ru{cool},}{\quad \text{for }~k=1,\ldots,N}
\end{mini!}

\noindent where:

\begin{itemize}
\item $T_k = [T_{k,1}\ru{zon}, T_{k,1}\ru{wall},\ldots, T_{k,M}\ru{zon}, T_{k,M}\ru{wall}]$ represents the state at time step $k$ for each of the $M$ scenarios.
\item $d_{k,i}=[T\ru{amb}_{k,i}, I_{k,i}, \theta\ru{occ}_{k,i}]$, represents the $i\ru{th}$ disturbance scenario at time step $k$.
\item The cost function is the average across all scenarios of the weighted average between the energy cost $J\ru{e}_k$ and the discomfort cost $J\ru{d}_k$ of each specific scenario.
\item The input control  $q_k = [q_k\ru{heat}, q_k\ru{cool}]$ remains equal across all scenarios.
\item The building dynamics are represented independently for each scenario by \eqref{eq:dyn2}.
\item  The constant parameters are the same as for deterministic MPC.
\end{itemize}

\begin{remark}
Note that other stochastic formulations exist that can provide guarantees on the feasibility of the obtained solution, e.g.\ \cite{ParFab:14}. In this paper, we adopt the scenario-based formulation (also referred to as \textit{multi-scenario} formulation) as in e.g.\cite{TiaNeg:17,TiaGuo:19,VelMae:16,VelMae:17}. The implementation of other stochastic methods will be investigated as future work.
\end{remark}

\subsection{Linear MPC}
\label{sec:linearMPC}
We will compare the SBMPC approach against two linear MPC approaches: deterministic linear MPC and linear SBMPC. In both cases, the optimization problems solved at each time step are the same as the ones defined by \eqref{eq:optidet} and \eqref{eq:optisb} but with a minor modification. Instead of using the nonlinear dynamics \eqref{eq:dyn} and \eqref{eq:dyn2}, the dynamics are given by the linear model defined in \eqref{eq:linearmodel}. In particular, for linear deterministic MPC, constraint \eqref{eq:dyn} is replaced by:

\begin{equation}
T_{k+1}= A\,T_{k} + B_1\,q_k + B_2\,d_{k},\quad \text{for }~k=1,\ldots,N.
\end{equation}

\noindent Similarly, for linear SBMPC, constraint \eqref{eq:dyn2} is replaced by:
\begin{align}
T_{k+1,i}= A\,T_{k,i} + B_1\,q_k + B_2\,d_{k,i},\quad &\text{for }~i=1,\ldots,M\nonumber\\ &\text{for }~k=1,\ldots,N.
\end{align}

\section{Scenario generation method}\label{sec:Scenarios}

\label{sec:forecasting}
In this section, we describe the scenario generation method for modeling the uncertainty in the system disturbances.

\subsection{Introduction}
Let us define  a random variable $X$ representing some time series process, e.g.\ external temperature, and the related multidimensional random variable $\mathbf{X}$ representing the distribution of $X$ in a time grid of $N$ time steps, i.e.\ $\mathbf{X}=[X_1,\ldots,X_N]\tr$. To generate scenarios, we will build the multivariate distribution of $\mathbf{X}$, i.e.\ $F(\mathbf{X})$, so that by sampling from $F(\mathbf{X})$ we can obtain $M$ scenarios of $\mathbf{X}$, i.e.\ $\mathbf{x}^1,\ldots,\mathbf{x}^M$.

When building $F(\mathbf{X})$, in order to satisfy the desired properties of scenario generation methods (see Section \ref{sec:introscen}), several requirements need to be satisfied:

\begin{itemize}
    \item $F(\mathbf{X})$ should not be substituted by the $N$ marginal distributions $F(X_1),\ldots,F(X_N)$. In particular, $F(X_i)$ only represents the distribution of $X$ at time step $i$ but does not consider the correlation between $X_1,\ldots,X_N$.
    \item $F(\mathbf{X})$ should not be built as a stationary distribution. Instead, the distribution should consider the properties of the underlying random variable $\mathbf{X}$. For example, in the case of temperature or solar irradiance, it is clear that $F(\mathbf{X})$ should vary with the day of the year $d$ as well as the hour of the day $h$, i.e.\ $F(\mathbf{X}):=g(\mathbf{X},d,h)$.
    \item The distribution $F(\mathbf{X})$ should include any external dependency of $\mathbf{X}$. For instance, if $\mathbf{X}$ represents the ambient temperature, $F(\mathbf{X})$ needs to explicitly include the dependency w.r.t.\ factors like the solar irradiance $I$, i.e.\ $F(\mathbf{X}):=g(\mathbf{X},I)$.
\end{itemize}

\subsection{Scenario generation method}
The proposed method consists of four steps: 
\begin{enumerate}
\item generation of a deterministic forecast $\bar{\mathbf{x}}$ of the random variable $\mathbf{X}$.
\item Generation of the marginal probability distribution  $F(X_1),\ldots,F(X_N)$ along the horizon $N$.
\item Generation of the distribution $F(\mathbf{X})$ using a parametric copula and the marginal distributions $F(X_1),\ldots,F(X_N)$.
\item Sampling of scenarios using $F(\mathbf{X})$.
\end{enumerate}
In this section, we explain the  four steps in detail.

\subsubsection{Deterministic forecast}
To build a deterministic/point forecast of the variable of interest we employ state-of-the-art methods for each variable of interest:

\begin{itemize}
\item For the solar irradiance, we consider two forecasting models: the deep neural network proposed in \cite{Lago2018b} for the short-term predictions (anything below 6 hours), and the ECMWF weather forecast \cite{ecmwf} for long-term predictions (anything beyond 6 hours). This distinction is made because, in the context of solar irradiance forecasting, machine learning techniques perform better for the short-term horizons, while numerical weather forecasts are more accurate for long-term horizons \cite{Lago2018b} (see also Remark \ref{rem:junction}). \blue{For what concerns the long-term prediction forecasts, note that weather-based models are highly complex models based on weather patterns; for practical applications, these forecasts are not produced by researcher but rather purchased from three main providers of weather forecasts worldwide and for our study we purchased them from ECMWF \cite{ecmwf}}.
\item For  the  ambient  temperature,  considering  the  recent success of deep learning methods for forecasting energy-related  variables  [66–72],  we  develop  a  deep neural network that uses as inputs the past values of the ambient temperature (hourly values over the last three days), the ECMWF weather forecast of the solar irradiance on hourly resolution over the forecasting horizon, and the hour of the day and day of the year when the prediction is made. For the deep neural network, we consider a two-hidden layer architecture whose parameters are optimized using hyperopt \cite{Bergstra2013}, a Bayesian optimization algorithm. We optimize the number of neurons per layer and the activation function. As a result of the optimization we considered a deep neural network with 240 (first hidden layer) and 135 (second hidden layer) neurons. The network uses the rectifier linear unit as activation function and is optimized using the Adam \cite{Kingma2014} optimizer with early stopping.
\end{itemize}

It is important to note that the other steps to generate scenarios are independent of the method employed to generate the deterministic forecast. As such, while we advocate for the use of state-of-the-art methods to obtain the most accurate scenarios, the proposed methodology would work as well with any deterministic forecast.

\blue{
\begin{remark}\label{rem:junction}
The splits between horizon regarding the forecast of the irradiance are a well studied problem in the literature (see e.g.\ \cite{Perez2010,Lago2018b}). Note that the exact split (4, 5, 6 hours) might be specific to the location.
\end{remark}
}

\subsubsection{Marginal distributions}
\label{sec:marginal}
To generate the marginal distributions, considering its simplicity yet high accuracy, we employ the method of empirical quantiles \cite{Nowotarski2018}. In detail, to generate the marginal distribution $F(X)$ of a variable $X$, the simplest version of this method consists of four steps:

\begin{enumerate}
    \item Consider deterministic forecasts of the variable in the past, e.g.\ $\bar{x}_{1}, \ldots, \bar{x}_{n}$. 
    \item Compute the associated historical forecasting errors of the deterministic forecast, e.g.\ $\bar{\epsilon}_{1}, \ldots, \bar{\epsilon}_{n}$. 
    \item Compute the empirical quantiles of the errors and its associated empirical distribution  $F(\epsilon)$.
    \item Model the marginal distributions as the point forecast plus the marginal distribution of the errors:
\begin{equation}
    F(X) = \bar{x} + F(\epsilon).
\end{equation}
\end{enumerate}

For the proposed approach, the method is modified in order to model non-stationary marginal distributions. In particular, defining $X_{k,h,d}$ as the random variable representing the value of $X$ at time $h$ of day $d$ that is predicted $k$ time steps ahead, the proposed approach estimates each distribution $F(X_{k,h,d})$ independently. To do so, the distributions of the errors $\epsilon_{k,h,d}$ are independently considered  for each time step $k$, time $h$, and day $d$, and the distribution $F(X_{k,h,d})$ is estimated accordingly:

\begin{equation}
    F(X_{k,h,d}) = \bar{x}_{k,h,d} + F(\epsilon_{k,h,d}).
\end{equation}

\noindent In addition, the following two considerations are made:

\begin{itemize}
\item To obtain non-stationary marginal distributions that explicitly model the variability of the distribution along a year, $F(\epsilon_{k,h,d})$ is estimated using the historical errors of the last 60 days.
    \item To explicitly model the variability of the distribution with the time step and time of the day, $F(\epsilon_{k,h,d})$ is estimated using past errors of the deterministic forecasts made for the same time step $k$ and time of the day $h$. Particularly, $F(\epsilon_{k,h,d})$ is estimated using the historical errors $\bar{\epsilon}_{k,h,d-1},\bar{\epsilon}_{k,h,d-2},\ldots,\bar{\epsilon}_{k,h,d-n}$.
\end{itemize}

\subsubsection{Scenario generation}\label{sec:ScenarioGeneration}
To generate scenarios, we consider the marginal distributions estimated in the previous step, a Gaussian copula function \cite{Pinson2009}, and Sklar's Theorem \cite{Sklar1959,Sklar1973}. In detail, let us define an $N$-dimensional random variable $\mathbf{X}=[X_1,\ldots,X_N]^\top$, its associated marginal distributions by $F_1(X_1),\ldots,F_N(X_N)$, and the multivariate cumulative distribution by $F(X_1,\ldots,X_N)$. If the marginals are continuous, Sklar's theorem states that there is a copula function $\mathcal{C}:[0,1]^N \rightarrow [0,1]$ such that:

\begin{equation}
    F(X_1,\ldots,X_N) = \mathcal{C} \bigl( F_1(X_1),\ldots,F_N(X_N) \bigr).
\end{equation}

\noindent In other words, assuming that the copula function is known, the multivariate cumulative distribution can be easily obtained if the marginal distributions are known.

Using this theorem, to generate scenarios, we employ one of the copulas functions that requires fewer computational time: the Gaussian copula. This selection is done for three reasons: i) the method to generate scenarios should be fast for real time implementation; ii) empirically, we observed the Gaussian copula to be a good fit for the disturbances considered, i.e.\ ambient temperature and irradiance; iii) the Gaussian copula is a well established method that has been used to generate scenarios for different energy-based applications  \cite{Pinson2009,Golestaneh2016}.

For the sake of simplicity, we refer to \cite{Pinson2009} for details on the estimation of the Gaussian copula. Here, we simply outline the main idea of the method, which relies on two random variables transformations:

\begin{enumerate}
    \item Given a marginal distribution $F_i(X_i)$ of a random variable $X_i$, we can define a new random variable $Y_i=F_i(X_i)$. Due to the properties of $F_i(X_i)$, it can be easily shown that $Y_i\sim\mathcal{U}[0,1]$, i.e.\ the new random variable follows an uniform distribution.
    \item Given a random variable $Y_i\sim\mathcal{U}[0,1]$, we can obtain a random variable $Z_i=\Phi(Y_i)\sim\mathcal{N}(0,1)$ that is normally distributed, where $\Phi$ is the probit function.
\end{enumerate}

\noindent Then, to generate $M$ scenarios of $\mathbf{X}$ at time step $k$, i.e.\ $\{\bar{\mathbf{x}}^j_k=[\bar{x}^j_{k,1},\ldots,\bar{x}^j_{k,N}]\tr\}_{j=1}^M$, the method consist of 5 steps:

\begin{enumerate}
    \item Consider historical realizations $\mathbf{x}_{k-1},\ldots,\mathbf{x}_{k-n}$ of $\mathbf{X}$.
    \item Use the marginal distributions $F_1(X_1),\ldots,F_N(X_N)$ to map each historical sample
    $\mathbf{x}_i=[x_{i,1},\ldots,x_{i,N}]\tr$ to a transformed sample $\mathbf{z}_i=[z_{i,1},\ldots,z_{i,N}]\tr$, where $Z_{i,j}\sim\mathcal{N}(0,1)$.
    \item Compute the covariance matrix $\Sigma$ of the historical transformed samples $\mathbf{z}_{k-1},\ldots,\mathbf{z}_{k-n}$.
    \item Draw $M$ samples $\bar{\mathbf{z}}^1,\ldots,\bar{\mathbf{z}}^M$ from the normal distribution $\mathcal{N}(0,\Sigma)$.
    \item Use the inverse of the two transformations applied in the previous steps to map the samples $\bar{\mathbf{z}}^1,\ldots,\bar{\mathbf{z}}^M$ to a set of samples $\bar{\mathbf{x}}^1,\ldots,\bar{\mathbf{x}}^M$.
\end{enumerate}

\noindent The samples $\bar{\mathbf{x}}^1,\ldots,\bar{\mathbf{x}}^M$ represent the $M$ required scenarios $\{\bar{\mathbf{x}}^j_k=[\bar{x}^j_{k,1},\ldots,\bar{x}^j_{k,N}]\tr\}_{j=1}^M$. In particular, they follow the original marginal distributions $F_1(X_1),\ldots,F_N(X_N)$ and they model the inter-correlation in $\mathbf{X}=[X_1,\ldots,X_N]^\top$.

As it was done for the marginal distributions, the Gaussian copula method is modified in order to model non-stationary distributions. In particular, defining $X_{k,h,d}$ as the random variable representing the value of $X$ at time $h$ of day $d$ and predicted with a time step $k$, the proposed approach estimates the copula function with the two following modifications:

\begin{itemize}
\item To have non-stationary distributions that explicitly model the variability of the distribution along a year, the copula function is estimated using the historical data of the last 60 days.
    \item To explicitly model the variability of the distribution with the time step and time of the day, the copulas are estimated using marginal distributions $F(\epsilon_{k,h,d})$ that explicitly model the distribution of $X$ as a function of the time step ahead $k$, time of the day $h$, and day of the year $d$.
\end{itemize}

\begin{figure}[t]
	\centering
	\includegraphics[width=0.48\textwidth]{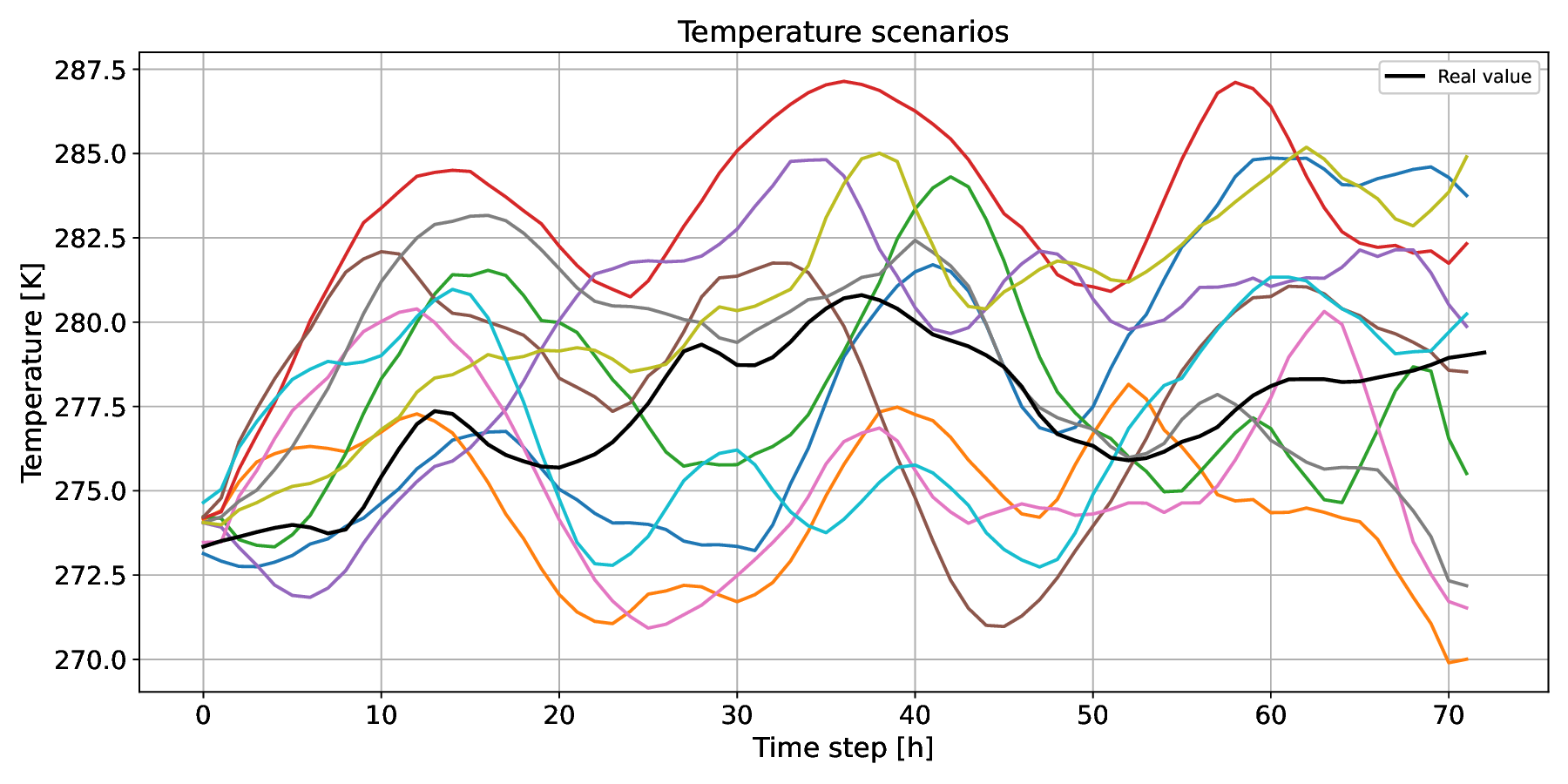}
	\caption{10 temperature scenarios obtained with the method presented in Section \ref{sec:Scenarios}. A time step corresponds one hour. The actual measured value of the temperature is shown in black color.}
	\label{fig:temperatureScenarios}
\end{figure}

\begin{figure}[t]
	\centering
	\includegraphics[width=0.48\textwidth]{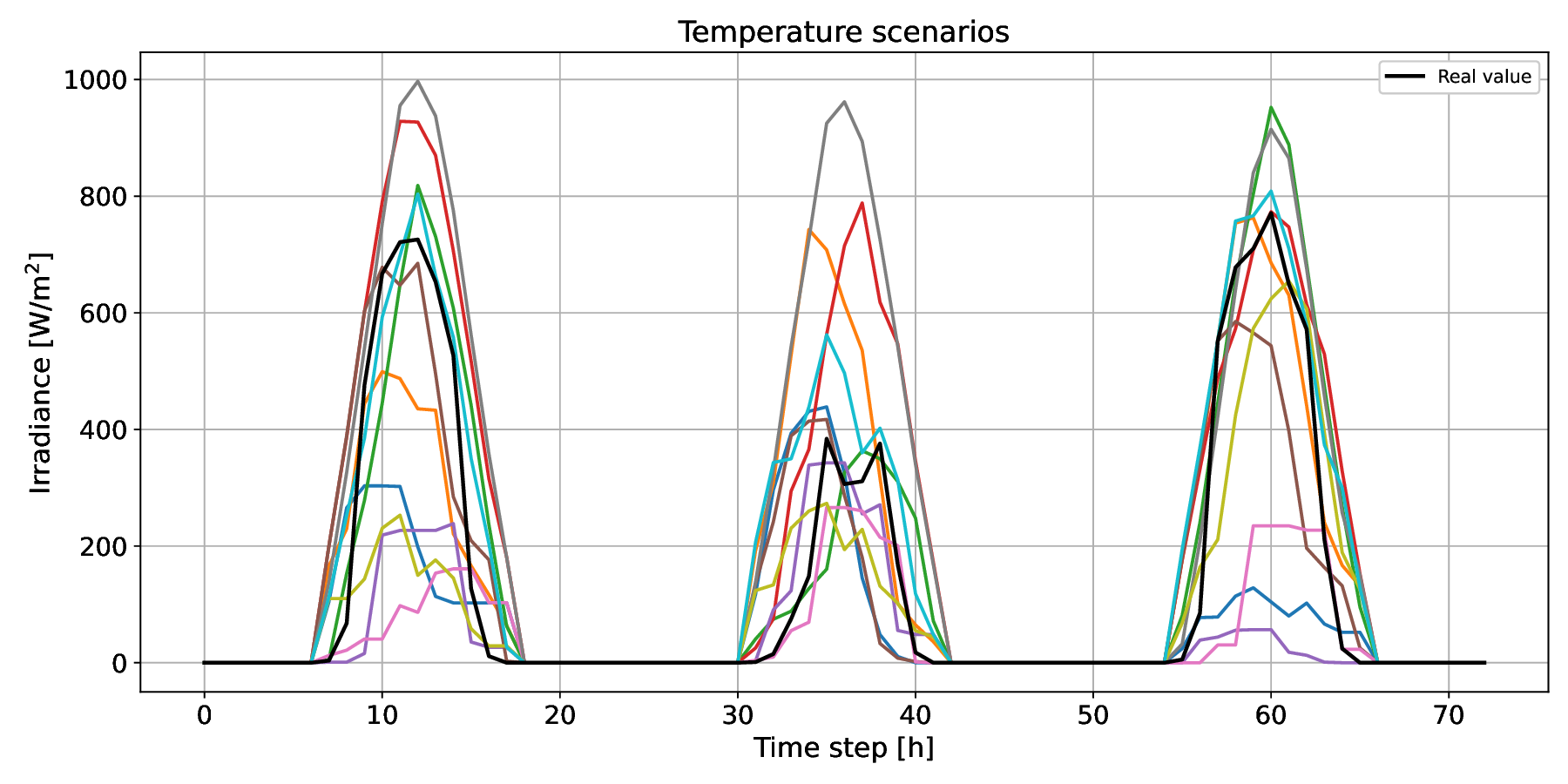}
	\caption{10 solar irradiance scenarios obtained with the method presented in Section \ref{sec:Scenarios}. A time step corresponds one hour. The actual measured value of the temperature is shown in black color.}
	\label{fig:irradianceScenarios}
\end{figure}

We show 10 temperature scenarios and 10 solar irradiance scenarios in Figures \ref{fig:temperatureScenarios} and \ref{fig:irradianceScenarios}, respectively.

\blue{To evaluate the marginal distributions, we compute their 90\% and 80\% coverage. That is, we compute the percentage of historical elements that do indeed fall in the interval [5\%, 95\%] and [10\%, 90\%]. As can be seen from Table \ref{tab:coverage}, the coverage of the distributions is acceptable. As could be expected, the distribution of the irradiance is more far off, but overall the coverage is within expected errors.}

\blue{
\begin{remark}
Let us explain here how we capture the correlation between the different variables. Since we generate scenarios using copula functions and given a set of individual variables with their marginal distribution, we can use a copula function to represent the full distribution of all the variables. For that, we use the individual marginal distributions and some map/inference process. The individual variables can be either the same variable at different time points, e.g.\ the temperature at different hours, or different variables altogether. From the perspective of the method it does not make any difference. In both cases we consider the marginal distributions of the individual variables (something that we can compute with historical data), and we map them to a copula function that captures the full distribution of the variables (including correlation).
\end{remark}
}

\begin{table}[h!]
\blue{
\begin{center}
\begin{tabular}{l|lll}
            & \multicolumn{2}{l}{Coverage} &  \\
            & 90\%      & 80\%       &  \\
            \hline
Temperature & 95.59\%   & 84.71\%         &  \\
Irradiance  & 85.56\%   & 75.85\%         & 
\end{tabular}
\caption{Coverage of the temperature and irradiance scenarios}
\label{tab:coverage}
\end{center}
}
\end{table}

\subsubsection{Properties}
As a final remark, we outline how the proposed method satisfies each of the required properties mentioned in Section \ref{sec:introscen}:

\begin{enumerate}
	\item As the distribution of the disturbances are modelled with non-parametric quantile functions, the generated scenarios are not restricted to the standard assumption of Gaussian forecasting errors.
	\item Since the scenarios are generated using a copula function, the multivariate distribution is explicitly considered and the scenarios include the time correlations.
	\item As the marginal distributions are estimated for each hour of the day, time of the year, and time-step, the resulting multivariate distribution is non-stationary and captures all time dependencies.
	\item As the point forecast considers external factors, the method is not limited to historical data of the variable of interest.
	\item Since the Gaussian copula and the empirical quantile methods have low computational costs, the method is especially suitable for online optimization.
	
\end{enumerate}

\begin{remark}
In this section, we presented the properties that scenario generation methods should have and we adopt the method presented in Section \ref{sec:ScenarioGeneration}, qualitatively comparing it against other scenario generation methods used in the literature for SBMPC. However, a comparative and quantitative study comprehending several scenario generation methods aimed at investigating which one provides the best control performance is out of scope for this work and is left as a suggestion for future work in Section \ref{sec:Conclusions}.
\end{remark}

\section{Case study}\label{sec:CaseStudy}
\begin{table}[!t]
	\begin{center}
		\def\arraystretch{1.3}
		\begin{tabular}{r|c|l}
			\textbf{Parameter} &  \textbf{Value}&\textbf{Definition} \\
			\hline
			$\overline{Q}\ru{heat}\rs{max}$ [W] & 500000& Maximum heating power	\\
			$\overline{Q}\ru{cool}\rs{max}$ [W] & 300000& Maximum cooling power \\
			  $\eta\ru{cool}$ &2.5 & Cooling efficiency \\
						 ${\eta}\ru{gas}$ & 0.9& Heating efficiency \\
			$c\rs{gas}$ [\euro{}/kWh] & 0.041& Gas cost \\
			$c\rs{ele}$ [\euro{}/kWh] & 0.15& Electricity cost\\
		\end{tabular}
	\end{center}
	\caption{Parameters of the building considered in Section \ref{sec:CaseStudy}}
	\label{tab:heatprod}
\end{table}

\begin{table*}[t]
\begin{center}
  \begin{tabular}{C{2.8cm}|C{2cm} |C{2cm} |C{2cm} |C{2cm}} 
 	 & $\alpha = 50$ & $\alpha = 100$ & $\alpha = 200$ & $\alpha = 500$ \\ 
 	\hline 
PIMPC & 8104 & 9009  & 10573  & 14676\\ \hline
DetMPC-Lin & 11770 & 17773 & 29722 & 65616 \\
DetMPC-Mod & 10462 & 12236 & 15308 & 23961 \\ \hline
SBMPC-Mod-10 & 8994 & 10590 & 14032 &	20603 \\ 
SBMPC-Mod-20 & 9909 & 10767 & 13778 & 21247 \\ 
SBMPC-Mod-30 & 9417 & 11088 & 13204 & 21466 \\ 
SBMPC-Mod-40 & 10517 & 11950 & 14014 &	20888 \\ \hline
SBMPC-Lin-10 & 8519 & 15856 & 19203 & 48135 \\ 
SBMPC-Lin-20 & 8810 & 13556 & 22498 & 51368 \\ 
SBMPC-Lin-30 & 11477 & 11485 & 23472 & 49714 \\ 
SBMPC-Lin-40 & 8485 & 12957 & 30431 &	33807
\end{tabular}
\vspace{6pt}
 \caption{Total closed-loop costs for all the controllers considered in the case study.}
 \label{tab:totalCosts}
\end{center}
\end{table*}

\begin{table*}[t]
\begin{center}
\blue{
  \begin{tabular}{C{2.8cm}|C{2.7cm} |C{2.7cm} |C{2.7cm} |C{2.7cm}} 
 	 & $\alpha = 50$ & $\alpha = 100$ & $\alpha = 200$ & $\alpha = 500$ \\ 
 	\hline 
PIMPC & 80.9$\%$ $\vert$ 19.1$\%$ & 74.8$\%$ $\vert$ 25.2$\%$ & 65.7$\%$ $\vert$ 35.3$\%$ & 49.8$\%$ $\vert$ 50.2$\%$\\ \hline
DetMPC-Lin & 31.0$\%$ $\vert$ 69.0$\%$ & 20.0$\%$ $\vert$ 80.0$\%$ & 11.7$\%$ $\vert$ 88.3$\%$ & \white{0}5.2$\%$ $\vert$ 94.8$\%$\\
DetMPC-Mod & 66.0$\%$ $\vert$ 34.0$\%$ & 58.1$\%$ $\vert$ 41.9$\%$ & 47.6$\%$ $\vert$ 52.4$\%$ & 31.1$\%$ $\vert$ 68.9$\%$\\ \hline
SBMPC-Mod-10 & 72.1$\%$ $\vert$ 27.9$\%$ & 66.0$\%$ $\vert$ 34.0$\%$ & 51.3$\%$ $\vert$ 48.7$\%$ & 35.5$\%$ $\vert$ 64.5$\%$\\
SBMPC-Mod-20 & 70.2$\%$ $\vert$ 29.8$\%$ & 66.1$\%$ $\vert$ 33.9$\%$ & 52.4$\%$ $\vert$ 47.6$\%$ & 34.5$\%$ $\vert$ 65.5$\%$\\
SBMPC-Mod-30 & 71.2$\%$ $\vert$ 28.8$\%$ & 62.7$\%$ $\vert$ 37.3$\%$ & 53.9$\%$ $\vert$ 46.1$\%$ & 34.7$\%$ $\vert$ 65.3$\%$\\
SBMPC-Mod-40 & 66.5$\%$ $\vert$ 33.5$\%$ & 59.9$\%$ $\vert$ 40.1$\%$ & 51.4$\%$ $\vert$ 48.6$\%$ & 36.6$\%$ $\vert$ 63.4$\%$\\ \hline
SBMPC-Lin-10 & 36.4$\%$ $\vert$ 63.6$\%$ & 22.3$\%$ $\vert$ 77.7$\%$ & 15.2$\%$ $\vert$ 84.8$\%$ & \white{0}6.8$\%$ $\vert$ 93.2$\%$\\
SBMPC-Lin-20 & 36.1$\%$ $\vert$ 63.9$\%$ & 23.6$\%$ $\vert$ 76.4$\%$ & 14.0$\%$ $\vert$ 86.0$\%$ & \white{0}6.6$\%$ $\vert$ 93.4$\%$\\
SBMPC-Lin-30 & 31.9$\%$ $\vert$ 68.1$\%$ & 26.0$\%$ $\vert$ 74.0$\%$ & 13.9$\%$ $\vert$ 86.1$\%$ & \white{0}6.7$\%$ $\vert$ 93.3$\%$\\
SBMPC-Lin-40 & 31.4$\%$ $\vert$ 68.6$\%$ & 23.7$\%$ $\vert$ 76.3$\%$ & 10.6$\%$ $\vert$ 89.4$\%$ & \white{0}8.3$\%$ $\vert$ 91.7$\%$\\
\end{tabular}
\vspace{6pt}
 \caption{\blue{Contribution of the subcosts $J\rs{e}$ and $\alpha J\rs{d}$ to the total closed-loop for all the controllers considered in the case study. The first number represents the percentage of the energy cost $J\rs{e}$ in the total cost, while the second number represents the percentage of the discomfort cost multiplied by $\alpha$, i.e.\ $\alpha J\rs{d}$. This table is also depicted as a stacked bar plot in Figure \ref{fig:percentageCost}.}}
 \label{tab:percentageCost}
 }
\end{center}
\end{table*}

\begin{table*}[t]
\begin{center}
  \begin{tabular}{C{3cm}|C{2cm} |C{2cm} |C{2cm} |C{2cm}} 
 	 & $\alpha = 50$ & $\alpha = 100$ & $\alpha = 200$ & $\alpha = 500$ \\ 
 	\hline 
PIMPC & 49.2 & 36.7 & 29.6 & 25.4 \\ \hline
DetMPC-Lin & 156.8 & 146.4 & 140.5 & 137.2 \\
DetMPC-Mod & 93.3 & 74.9 & 62.7 & 53.7 \\ \hline
SBMPC-Mod-10 & 69.0 & 58.0 & 51.6 & 41.9 \\ 
SBMPC-Mod-20 & 79.7 & 53.2 & 49.4 & 43.5 \\ 
SBMPC-Mod-30 & 71.7 & 60.8 & 50.0 & 42.4  \\ 
SBMPC-Mod-40 & 90.8 & 67.2 & 50.6 & 41.7  \\ \hline
SBMPC-Lin-10 & 115.6 & 129.9 & 90.9 & 105.9 \\ 
SBMPC-Lin-20 & 100.8 & 114.7 & 103.3 & 112.9  \\
SBMPC-Lin-30 & 147.7 & 98.8 & 115.3 & 111.6 \\ 
SBMPC-Lin-40 & 124.2 & 105.2 & 113.7 & 83.4
\end{tabular}
\vspace{6pt}
 \caption{Total amount of discomfort measured in $\mathrm{Kh}$.}
 \label{tab:Kh}
\end{center}
\end{table*}

\begin{figure*}[h!]
	\centering
	\includegraphics[width=0.95\textwidth]{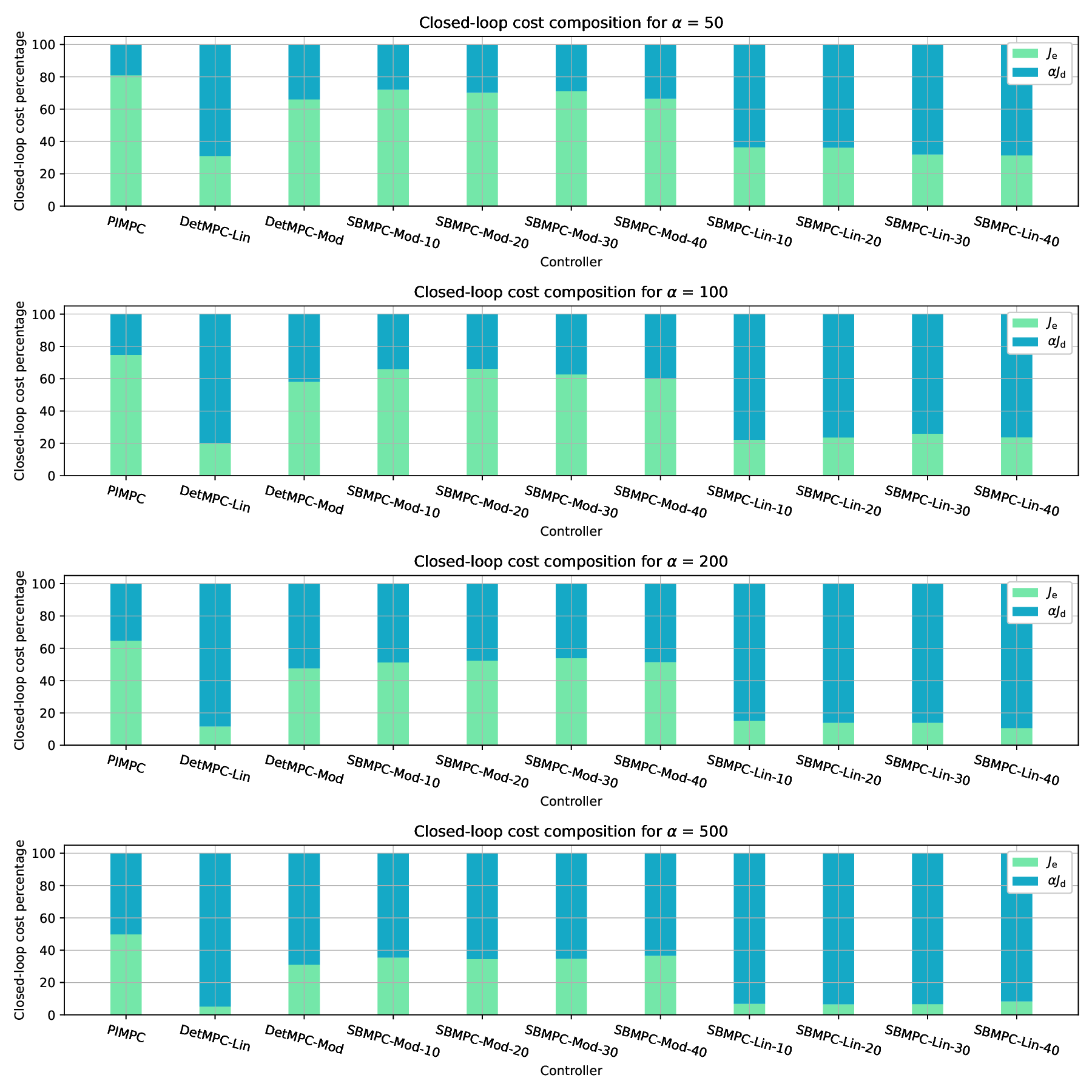}
	\caption{\blue{Contribution of the subcosts $J\rs{e}$ and $J\rs{d}$ to the total closed-loop for all the controllers considered in the case study, for each different controller and value of $\alpha$. The data used for this plot is shown in Table \ref{tab:percentageCost}.}}
	\label{fig:percentageCost}
\end{figure*} 

We present in this section the simulation results in which we compare 5 different controllers:
\begin{itemize}
\item PIMPC: perfect-information MPC, obtained using the values of the measurements of the disturbances as if they were known in advance. It is of course not possible to have the real values of the actual measurements beforehand in practice, but this controller can be used as a benchmark for the ideal theoretical achievable performance.
\item DetMPC-Mod: deterministic MPC controller presented in Section \ref{sec:DetMPC} together with the nonlinear Modelica model.
\item SBMPC-Mod: SBMPC controller presented in Section \ref{sec:SBMPC} together with the nonlinear Modelica model.
\item DetMPC-Lin: deterministic MPC controller together with the linearized model presented in Section \ref{sec:linearization}.
\item SBMPC-Lin: SBMPC controller together with the linearized model.
\end{itemize}

First, the simulation setup is discussed, then the results and the discussions are presented.

\subsection{Setup}
The closed-loop control is applied as explained in Section \ref{sec:implementation}, i.e.\ the MPC problem is solved and the first input is applied to the system. Then, for all the controllers, the evolution of the real building between sampling times is simulated through Modelica.

\begin{figure}[t]
	\centering
	\includegraphics[scale=0.43]{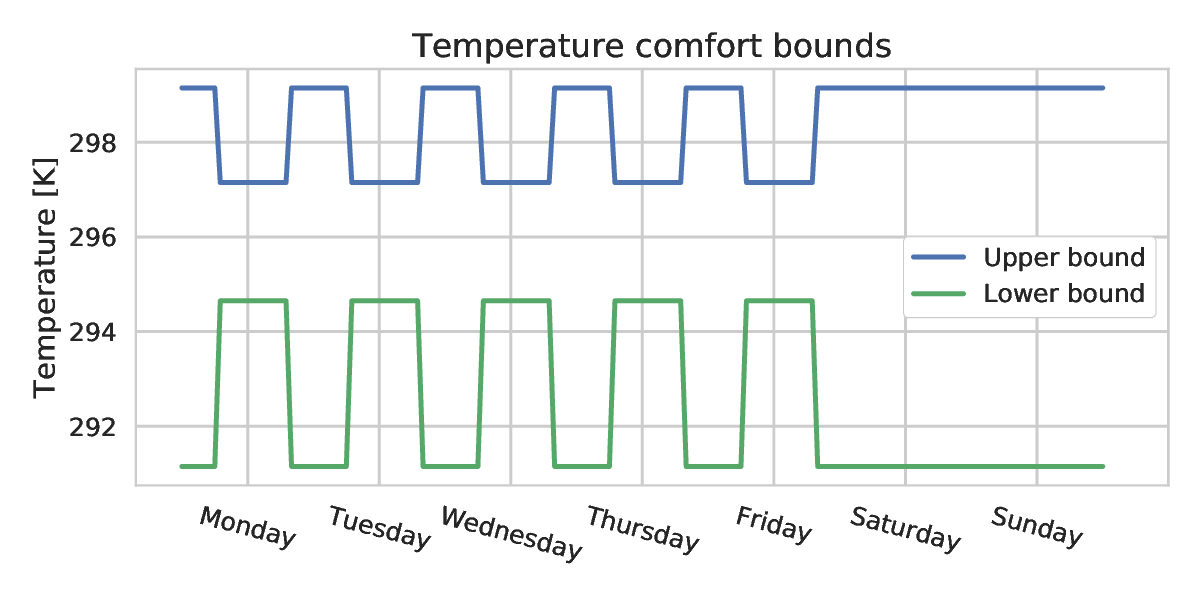}
	\caption{Comfort bound profiles.
	}
	\label{fig:occupation_hours}
\end{figure}

We perform simulations for one month in the winter season with $1\mathrm{h}$ sampling time, i.e.\ we solve $24\cdot 30 = 720$ optimization problems for each controller. The prediction horizon is $N\rl{p}=24$, i.e.\ corresponding to one day.
We consider an office building in Brussels, Belgium, with 7 floors and a total surface of 10000 $\text{m}^2$.  A nonlinear model of the building is estimated using Modelica based on the considerations of Section \ref{sec:officebuilding} and using data from the real building. In addition, a linear counterpart is also estimated using regular linear least squares as explained in Section \ref{sec:linearization}. The heating system consists of 2 gas boilers of $500$ kW each and one chiller of $500$ kW. We consider thermal comfort bounds that change throughout time, i.e.\ the lower and upper comfort bounds are set respectively to 21.5\degree C and 24\degree C during occupation hours and 18\degree C and 26\degree C during the non-occupation hours, as shown in Figure \ref{fig:occupation_hours}. Furthermore, the building occupancy profile follows the temperature comfort bounds, i.e.\ the occupancy is set to 1 when the comfort bounds are tight and 0 when they are loose. The solver and software tools used to solve the optimization problem are as explained in Section \ref{sec:implementation}. Lastly, the parameters of the building presented in Section \ref{sec:Control} are shown in Table \ref{tab:heatprod}.

For what concerns the SBMPC controllers, we choose 4 different number of scenarios: 10, 20, 30, and 40. Moreover, we perform the same simulations varying the parameter $\alpha$ in Section \ref{sec:Control}, choosing the values in the set $\{50,100,200,500\}$; recall that a higher $\alpha$ means a higher focus on the comfort of the occupants rather than on the economical cost. We indicate respectively by SBMPC-Mod-$n\rs{s}$ and by SBMPC-Lin-$n\rs{s}$, $n\rs{s}\in\{10,20,30,40\}$, the SBMPC controller with the Modelica model and the SBMPC controller with the linear model, considering $n\rs{s}$ scenarios.

\subsection{Results and discussion}
We focus our attention on four different aspects:
\begin{enumerate}
\item an analysis for the performance of the controller with different values of $\alpha$.
\item A comparison between the nonlinear Modelica model and the linear model.
\item A comparison between SBMPC strategies and DetMPC strategies.
\item A comparison between the SBMPC strategies with different numbers of scenarios. 
\end{enumerate}
Lastly, we pick a single representative optimization result and discuss it in more detail. 

We show the results of the simulations in Figures \ref{fig:percentageCost}, \ref{fig:comfortViolation}--\ref{fig:temperatureIrradiance} and Tables \ref{tab:totalCosts}--\ref{tab:Kh}. 
In Table \ref{tab:totalCosts}, we show the total closed-loop costs for each strategy and each different $\alpha$. \blue{In Table \ref{tab:percentageCost}, we show the percentage of each subcost with respect to the total closed-loop cost for each strategy and each different $\alpha$, i.e.\ we show $\frac{J\rs{e}}{J\rs{e}+\alpha J\rs{d}}$ and $\frac{\alpha J\rs{d}}{J\rs{e}+\alpha J\rs{d}}$.}
In Table \ref{tab:Kh} we show the total amount of discomfort using the unit measure $\mathrm{K\cdot h}$, as standard in the literature \cite{ParMol:13,Zhang2013,DeCHel:16,OldPar:12}, i.e.\ we show the integral of the comfort bounds violation. \blue{Figure \ref{fig:percentageCost} reports the results presented in Table \ref{tab:percentageCost} in a graphic way; the same holds for Figure \ref{fig:comfortViolation} and Table \ref{tab:Kh}.} Lastly, Figures \ref{fig:temperatureEvolution}--\ref{fig:temperatureIrradiance} show respectively the temperature evolution, the heating power, and the temperature and irradiance profiles for the representative simulation.

\subsubsection{Performance with different values of $\alpha$}\label{sec:alphaDiscussion}
In this section, we analyze how the different values of $\alpha$ alters the performance of the controllers in terms of energy cost and discomfort. This analysis should always be carried out when considering a case study as the one presented here, in order to understand which is the range of values for $\alpha$ that provide the best trade-off between comfort and energy cost reduction. 

From Tables \ref{tab:totalCosts}-\ref{tab:Kh} we can notice that the larger the $\alpha$, the larger, in general, the total costs and the lower the discomfort. This is as expected, since the role of $\alpha$ is to penalize the discomfort and a larger value means that we aim for a lower discomfort. We have noticed through simulations that, for this specific case study, a value of $\alpha$ lower than $50$ yields a very high and unacceptable discomfort cost, while for values larger than $500$, the energy costs increase highly without yielding a high reduction in the discomfort costs. Therefore, we focus our analysis on $\alpha$ in the range $\left[50,500\right]$.

We can observe that for $\alpha=50$ the discomfort cost is high and that the comfort could be improved by increasing $\alpha$. For $\alpha\geq 100$, the discomfort reaches more acceptable levels and this happens by consuming a larger quantity of energy in heating and thus increasing the total costs. However, the small decrease in the discomfort between the case $\alpha=500$ and $\alpha\in\{100,200\}$ does not seem to justify the large increase in the total cost observed for $\alpha=500$. Therefore, we can claim that, for this case study, the optimal values for $\alpha$ are in the range $\left[100,200\right]$. \blue{Note also that from Figure \ref{fig:percentageCost} and Table \ref{tab:percentageCost} we can notice how the two costs, i.e.\ energy and comfort, compose the total cost. As $\alpha$ increases, we notice an increase in the $\alpha J\rs{d}$ cost with respect to the $J\rs{e}$, as a higher $\alpha$ penalizes more even the small deviations from the comfort bounds. Therefore, we should not wrongly conclude that less energy is consumed for higher values of $\alpha$, but rather that small deviations are penalized more.}

\subsubsection{Comparison between the nonlinear Modelica model and the linear model}\label{sec:modelicaVSlinear}
\blue{Recalling Remarks \ref{rem:linearModel}-\ref{rem:nonlinearModelTested}, we present here a comparison between the controllers that use the Modelica model with respect to the ones that use the linear model. While using the Modelica model in the closed-loop for computing the evolution of the system results in a bias towards the nonlinear controllers, it is nevertheless of interest analyzing how good the linear controllers are compared to the ones using the Modelica model.}

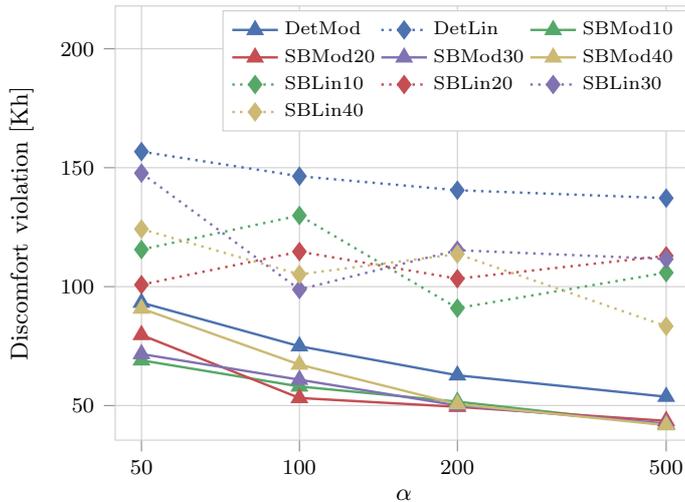
\begin{figure}
	\setlength{\figW}{0.5\textwidth}
	\setlength{\figH}{0.8\figW}
	\begin{center}
%
%
%
\begin{tikzpicture}

\definecolor{color1}{rgb}{0.333333333333333,0.658823529411765,0.407843137254902}
\definecolor{color0}{rgb}{0.298039215686275,0.447058823529412,0.690196078431373}
\definecolor{color3}{rgb}{0.505882352941176,0.447058823529412,0.698039215686274}
\definecolor{color2}{rgb}{0.768627450980392,0.305882352941176,0.32156862745098}
\definecolor{color5}{rgb}{0.392156862745098,0.709803921568627,0.803921568627451}
\definecolor{color4}{rgb}{0.8,0.725490196078431,0.454901960784314}

\begin{axis}[
xlabel={$\alpha$},
ylabel={Discomfort violation [Kh]},
xmin=44.5625469066872, xmax=561.009227150981,
ymin=35.4335080398767, ymax=218,
xmode=log,
width=\figW,
height=\figH,
tick align=outside,
xtick pos=left,
ytick pos=left,
xmajorgrids,
x grid style={white!80.0!black},
ymajorgrids,
y grid style={white!80.0!black},
axis line style={white!80.0!black},
legend cell align={left},
legend style={at={(0.99,0.99)}, anchor=north east, font=\scriptsize, draw=white!80.0!black},
legend columns=3, 
ylabel style={font=\small},
xlabel style={font=\small},
xticklabel style={font=\footnotesize},
yticklabel style={font=\footnotesize},
log ticks with fixed point,
x tick label style={/pgf/number format/1000 sep=\,},
xtick={10,50,100,200,500},
legend entries={{DetMod},{DetLin},{SBMod10},{SBMod20},{SBMod30},{SBMod40},{SBLin10},{SBLin20},{SBLin30},{SBLin40}
}
]
\addplot [line width=0.91pt, color0, mark=triangle*, mark size=3]
table {%
50 93.2639985759827
100 74.9313058019298
200 62.7108397772636
500 53.7013355032214
};

\addplot [line width=0.91pt, color0, dotted, mark options=solid,mark=diamond*, mark size=3]
table {%
50 156.769145837136
100 146.428088471368
200 140.536552417971
500 137.216675886822
};
\addplot [line width=0.91pt, color1, mark=triangle*, mark size=3]
table {%
50 69.0086863663962
100 58.0136813756961
200 51.6153038735958
500 41.8941809315966
};
\addplot [line width=0.91pt, color2, mark=triangle*, mark size=3]
table {%
50 79.6694045977958
100 53.1691470758961
200 49.4495503492159
500 43.5427898533967
};
\addplot [line width=0.91pt, color3, mark=triangle*, mark size=3]
table {%
50 71.650732946896
100 60.8555646598965
200 50.0331782529963
500 42.3993568797968
};
\addplot [line width=0.91pt, color4, mark=triangle*, mark size=3]
table {%
50 90.771299234996
100 67.2163660343955
200 50.5946949769965
500 41.7326371839966
};
\addplot [line width=0.91pt, color1, dotted, mark options=solid,mark=diamond*, mark size=3]
table {%
50 115.577249287495
100 129.917296004895
200 90.9464294034964
500 105.870518902896
};
\addplot [line width=0.91pt, color2, dotted, mark options=solid,mark=diamond*, mark size=3]
table {%
50 100.792092416996
100 114.720948711196
200 103.331140002088
500 112.939453001796
};
\addplot [line width=0.91pt, color3, dotted, mark options=solid,mark=diamond*, mark size=3]
table {%
50 147.735023264795
100 98.792769738896
200 115.323703974996
500 111.586845484595
};
\addplot [line width=0.91pt, color4, dotted, mark options=solid,mark=diamond*, mark size=3]
table {%
50 124.203741588996
100 105.156246993996
200 113.704408810996
500 83.3613546953966
};
\end{axis}

\end{tikzpicture}
	\end{center}
	\caption{Discomfort for the different controllers as a function of $\alpha$. The $x$-axis is in log-scale.}
	\label{fig:comfortViolation}
\end{figure}

\blue{From Table \ref{tab:totalCosts}, we can} see that all the controllers that use Modelica perform better than their linear counterparts for all the values of $\alpha\geq 100$. For $\alpha =50$, instead, the linear model yields a lower total cost than the one of Modelica in 3 out of 5 cases. Nevertheless, the linear model might seem to work better, i.e.\ to have a lower total cost, for a lower value of $\alpha$ because it always allows a large discomfort cost and it cannot manage well to keep the temperature within or close to the comfort bounds. The total cost can therefore be lower than for the Modelica-based controllers, but this occurs because the energy cost is low and the discomfort cost, although high, does not have a large impact on the total cost for a small $\alpha$. This also explains why for a large $\alpha$, i.e. for $\alpha\geq 100$, the total cost of the linear model controllers can become much higher than the one of the controllers that use Modelica. Indeed, the discomfort cost is always high, but the larger penalization, i.e.\ the larger $\alpha$, makes the total cost much higher. \blue{This fact can also be observed from Table \ref{tab:Kh}, where we observe a much higher discomfort for all the controllers with the linear models. The same conclusions can be drawn by analyzing Figure \ref{fig:comfortViolation}, which displays the results of Table \ref{tab:Kh}. Moreover, from Figure \ref{fig:percentageCost} we can notice how for all the values of $\alpha$ the controllers that use the linear model have a much higher comfort cost component than an energy cost component. The linear model does not employ more energy to reduce the comfort, which in turn results in high comfort violations. This concept will be discussed also in Section \ref{sec:representativeSim}.} \blue{We can therefore conclude that, independently of the bias mentioned in Remark \ref{rem:linearModel}, the linear model in this case fails to provide satisfactory performance in terms of comfort for the occupants of the building.}


\begin{figure}[t]
	\centering
	\includegraphics[width=0.48\textwidth]{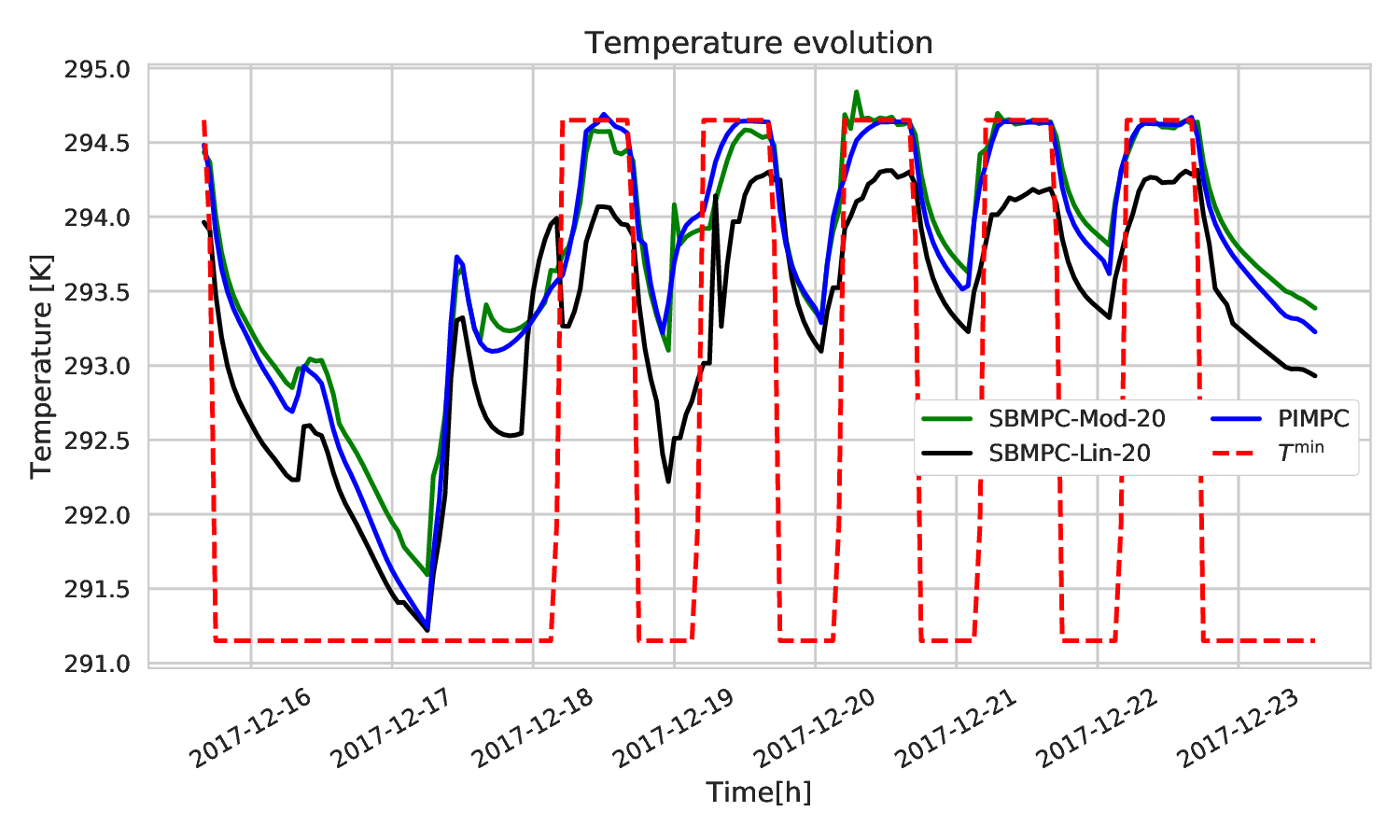}
	\caption{Temperature evolution during one week of a representative simulation with $\alpha=100$ and 20 scenarios for the SBMPC controllers.}
	\label{fig:temperatureEvolution}
	\includegraphics[width=0.48\textwidth]{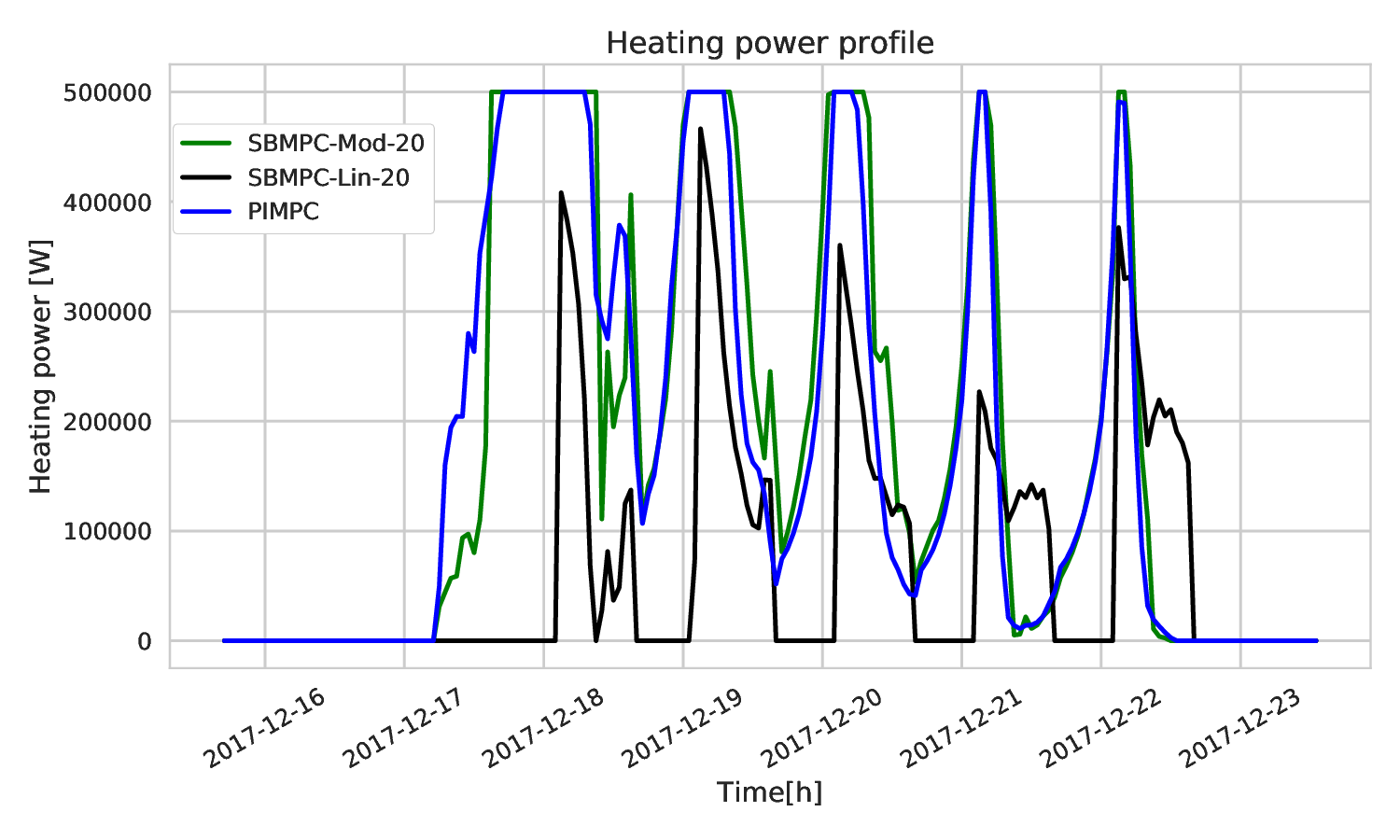}
	\caption{Heating power during 1 week of a representative simulation with $\alpha=100$ and 20 scenarios for the SBMPC controllers.}
	\label{fig:qHeaPro}
\end{figure} 

\begin{figure}[t]
	\centering
	\includegraphics[width=0.48\textwidth]{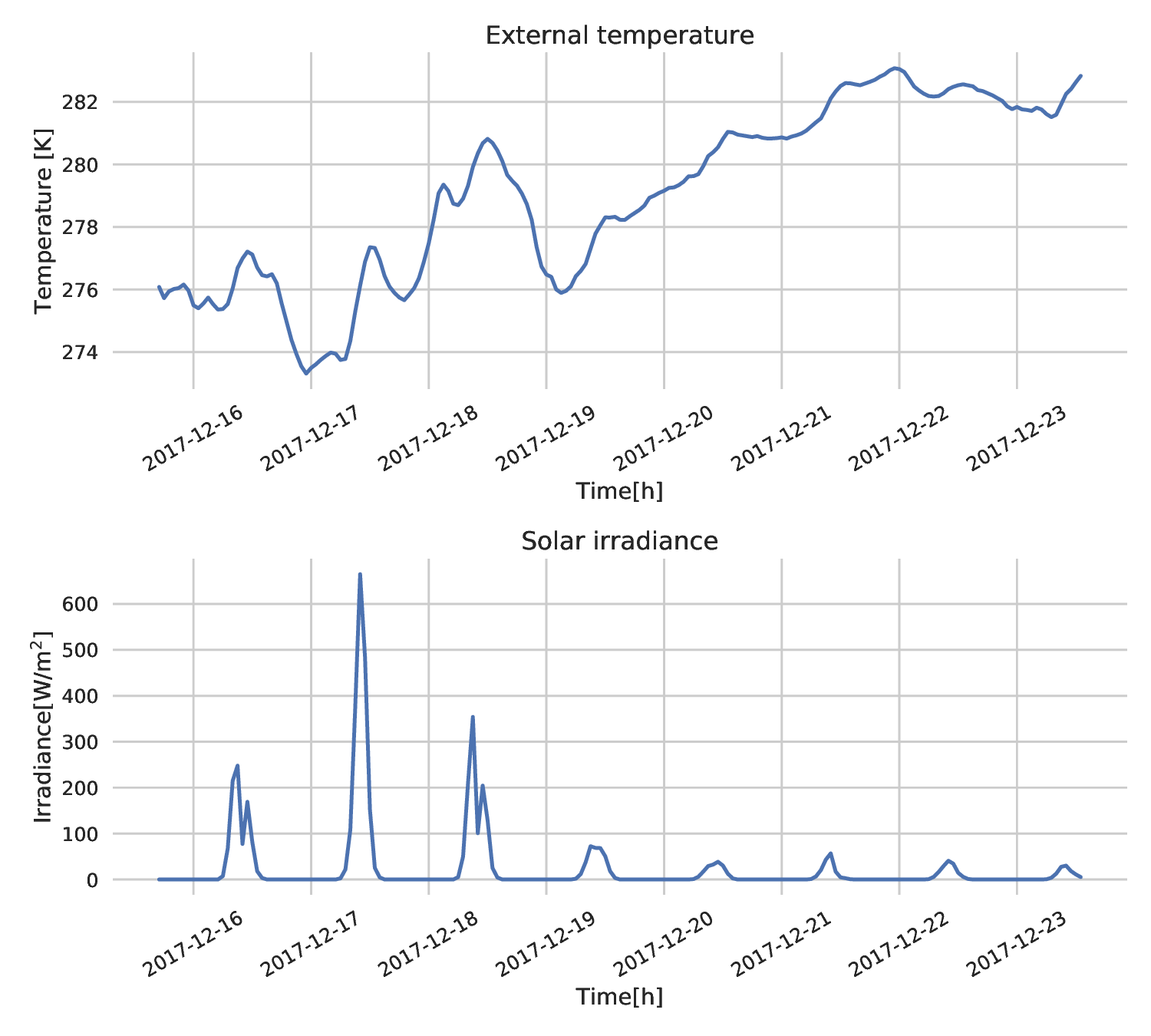}
	\caption{External temperature and solar irradiance during 1 week of a representative simulation with $\alpha=100$ and 20 scenarios for the SBMPC controllers.}
	\label{fig:temperatureIrradiance}
\end{figure}

\subsubsection{Comparison between SBMPC strategies and DetMPC strategies}
By checking again Table \ref{tab:totalCosts} we can compare the SBMPC strategies to the DetMPC ones. It can be noted from the table that, for all the values of $\alpha$, the SBMPC controllers perform almost always better than their deterministic counterpart, both for the linear and the nonlinear Modelica model. Note that the reduction, although not very large, is still consistent and it ranges from a minimum of $1.98\%$ to a maximum of $14.03\%$ for the controllers that use the Modelica model. Furthermore, by checking Table \ref{tab:Kh} and Figure \ref{fig:comfortViolation}, we can notice that the SBMPC strategies perform better than the DetMPC ones also in terms of comfort. Therefore, the SBMPC controllers can improve the overall performance, both in terms of total costs and discomfort, with respect to their DetMPC counterparts.

\subsubsection{Comparison between the SBMPC strategies with different number of scenarios} 
By analyzing the results of Table \ref{tab:totalCosts}, there does not seem to be a value for the number of scenarios that outperforms the other values, i.e.\ the performance does not seem to increase by increasing the number of scenarios. In 2 out of 4 columns of Table \ref{tab:totalCosts}, the SBMPC-Mod-20 achieves the best performance among the SBMPC-Mod controllers and in the other two cases, a number of scenarios equal to respectively 10 and 40 appear to be better than the other values. Therefore, increasing the number of scenarios does not seem to directly lead to a decrease in the total cost. This could be related to the fact that, while increasing the number of scenarios makes the system more robust to disturbances, it also makes the problem more complex to solve. Therefore, it might happen that, the larger the number of scenarios, more local minima exists, and the more likely it is that the solver converges to a suboptimal local minimum. \blue{Note that this issue does not affect the controllers with the linear model, as the problem solved in that case is a quadratic programming one; thus it is convex, and does not suffer from local minima issues.}

\subsubsection{Representative simulation}\label{sec:representativeSim}
In this section, we present a representative simulation, i.e.\ one week of simulation of the building, by showing the temperature evolution, heating power, and external disturbances for a specific value of $\alpha$ and of the number of scenarios. While the analysis of the results shown in this section are related to a specific case, the results can be generalized and the analysis of a case with a different $\alpha$ would be similar to what is here presented.

We compare three different control strategies here, namely SBMPC-Mod, SBMPC-Lin and PIMPC. We show in Figure \ref{fig:temperatureEvolution} the temperature evolution inside the room, in Figure \ref{fig:qHeaPro} the heating power, and in Figure \ref{fig:temperatureIrradiance} the temperature and irradiance profile, for one week of simulation with 20 scenarios and $\alpha=100$. For Figure \ref{fig:temperatureEvolution}, we also show the lower comfort bounds.

By analyzing the Figures \ref{fig:temperatureEvolution}--\ref{fig:qHeaPro}, we can note that the PIMPC manages to keep the temperature within the comfort bounds by using properly the heating power, thanks to the knowledge of the actual values of the future disturbances. For what concerns SBMPC-Mod and SBMPC-Lin, we can notice in Figure \ref{fig:temperatureEvolution} what we have already underlined in Section \ref{sec:modelicaVSlinear}, i.e.\ the fact that a controller that uses a linear model is not able to keep the temperature within the comfort bounds. We see indeed that, for most of the time, SBMPC-Lin yields the temperature profile that has the lowest value. This can also be observed from Figure \ref{fig:qHeaPro}, where we can notice that SBMPC-Lin heats less than SBMPC-Mod and it also starts heating later. On the other hand, SBMPC-Mod is able to maintain a larger temperature in the room and to be closer to the temperature comfort bounds. It uses more heating power, as can be seen from Figure \ref{fig:qHeaPro}, which leads to a higher energy cost compared to SBMPC-Lin, but also leads to an overall lower total cost and higher comfort from the user, as can be observed from Tables \ref{tab:totalCosts}--\ref{tab:Kh}.

In the first two days of simulation, which correspond to the weekend days, the heating power is turned off due to the low comfort bounds and to the prediction horizon of 24 h. However, it is possible to observe an increase in the room temperature, both for the day 16/12/2017 and 17/12/2017; for the latter, the rise in the temperature is even steeper. This can be explained by looking at Figure \ref{fig:temperatureIrradiance}, where we can observe that the day 17/02/2017 was a particularly sunny day. Hence, the steep increase in the temperature is due to large value of the solar irradiance in that particular day. This shows how important the influence of the external disturbances on the building can be, which once again underlines the importance of having good forecasts and it corroborates our choice of a scenario-based approach.

Note also that, while a longer prediction horizon could have been more beneficial and could have led to a higher performance of the controllers, we chose to focus our analysis on a horizon of 24 h as it provides a good trade-off between performance and computation time of all the simulations. The analysis performed here is still valid even when considering other prediction horizons. Moreover, carrying out a study on which prediction horizon is more beneficial for the considered building is beyond the scope of this paper.

\subsection{Summary}
We can summarize the observations obtained from the results of the simulations:
\begin{itemize}
    \item too low values of $\alpha$, i.e.\ $\alpha<100$, yield a high discomfort and too high values of $\alpha$, i.e.\ $\alpha\geq500$ yield a very large total cost without leading to a large improvement of the comfort. A trade-off between the two costs seems to be well achieved by a value of $\alpha$ between these two extrema, i.e.\ $\alpha\in\left[100,200\right]$. Moreover, tuning the parameter $\alpha$ is of great importance in order not to obtain a very large discomfort for the occupants of the building.
    \item \blue{For all the values of $\alpha$, the linear model shows a very large value of discomfort. In general, the linear model fails to capture many model dynamics, by heating much less and thus leading to temperatures that are well outside the comfort bounds.} 
    \item SBMPC can improve the performance with respect to DetMPC strategies, for both the linear and the Modelica models. This is due to the fact that SBMPC strategies consider different external disturbances scenarios and they have a large impact on the temperature evolution of the room.
    \item Increasing the number of scenarios does not seem to lead to a large decrease in the cost. This can be related to the fact that increasing the number of scenarios also increases the optimization complexity.
\end{itemize}

\blue{
\begin{remark}
Note that all the simulations performed in this work refer to a single, specific building. However, the controller designed can be applied to any building, as long as the model is adapted to the specific building under control. For instance, a similar controller, using a Modelica model and a deterministic MPC algorithm, has been applied to a different building with successful results in \cite{DeCHel:16}. Moreover, the scenario generation method presented here can also be applied to several disturbances that affect buildings. Therefore, while the results presented in this section refer to a single building, the method can be applied to any building. The results presented in Tables \ref{tab:totalCosts}--\ref{tab:Kh} will change if another building is considered, but we expect in any case similar results when applying the presented method to other buildings.
\end{remark}
}

\section{Conclusions}\label{sec:Conclusions}
We have presented a stochastic SBMPC controller using a Modelica nonlinear model that can be applied to building heating in buildings and that overcomes the limitations of both deterministic and linear MPC approaches. The building under control is affected by several external disturbances, e.g.\ outside temperature, solar irradiance, and we have proposed a new approach for generating disturbance scenarios that, unlike the existing methods from the literature, satisfies all the important properties of scenario generation methods for time series data. This proposed scenario generation method can be used in the SBMPC controllers.

To analyze and study the control approach, we considered a real building and performed several simulations to compare the controller that uses the linearized model against the controller that uses the Modelica model, different cost weights in the MPC cost function, deterministic MPC against SBMPC, and lastly different number of scenarios. Based on the results, we showed that SBMPC \blue{controllers outperform the deterministic MPC controllers both with and without the Modelica model}. \blue{At the same time, the linear model has shown not to be able to capture many model dynamics and this leads to poor performance, justifying the usage of more advanced models, e.g.\ Modelica ones}. 

As future work, \blue{we will develop a model with e.g.\ EnergyPlus to be used in the closed-loop simulations to improve the comparison performed between the linear and nonlinear model in Section 5. After such step, we will also perform experiments in the real building}. Moreover, a quantitative study on different scenario generation methods can be considered, in order to assess how the performance of the controller varies with the different scenario generation methods, as well as other stochastic MPC algorithms. \blue{Furthermore, a thorough comparison could be performed between our proposed method and other ones present in the literature of building heating systems, e.g.\ $\mathcal{H}_\infty$, fuzzy control, rule-based control. On top of that, a distributed or decentralized MPC controller can be developed to control independently each room, which might be more beneficial for very large buildings compared to controlling all the rooms with a single centralized controller.} Lastly, \blue{it is known that occupancy can largely affect the energy performance of buildings. Occupancy data was not available in this study, but we suggest to carry out a characterization study of such disturbance, so that occupancy scenarios can be generated and the overall performance of the proposed SBMPC controllers can be further assessed}.

\section*{Acknowledgments}
This research has received funding from the European Union’s Horizon 2020 research and innovation program under the Marie Sk\l odowska-Curie grant agreement No 675318 (INCITE). 

\bibliography{bibtex/bibtex,bibtex/allarticles}

\end{document}